\def\year{2022}\relax
\documentclass[letterpaper]{article} % DO NOT CHANGE THIS
\usepackage{aaai22}  % DO NOT CHANGE THIS
\usepackage{times}  % DO NOT CHANGE THIS
\usepackage{helvet}  % DO NOT CHANGE THIS
\usepackage{courier}  % DO NOT CHANGE THIS
\usepackage[hyphens]{url}  % DO NOT CHANGE THIS
\usepackage{graphicx} % DO NOT CHANGE THIS
\usepackage{natbib}  % DO NOT CHANGE THIS AND DO NOT ADD ANY OPTIONS TO IT
\usepackage{caption} % DO NOT CHANGE THIS AND DO NOT ADD ANY OPTIONS TO IT
\DeclareCaptionStyle{ruled}{labelfont=normalfont,labelsep=colon,strut=off} % DO NOT CHANGE THIS
\usepackage{algorithm,amsmath}
\usepackage{algorithmic, bm, amssymb, multirow}

\usepackage{newfloat}
\usepackage{listings}
\floatstyle{ruled}
\newfloat{listing}{tb}{lst}{}
\floatname{listing}{Listing}

\title{Scalable Quantum Architecture Search via Landscape Analysis}
\author{
    Chenghong Zhu\textsuperscript{\rm 1}\equalcontrib,
    Xian Wu\textsuperscript{\rm 1}\equalcontrib,
    Hao-Kai Zhang\textsuperscript{\rm 2},
    Sixuan Wu\textsuperscript{\rm 1},
    Guangxi Li\textsuperscript{\rm 3,}\textsuperscript{\rm 1},
    Xin Wang\textsuperscript{\rm 1}$^{\dagger}$
}
\affiliations{
    \textsuperscript{\rm 1} Thrust of Artificial Intelligence, Information Hub,\par The Hong Kong University of Science and Technology (Guangzhou), Guangdong 511453, China\\
    \textsuperscript{\rm 2} Institute for Advanced Study, Tsinghua University, Beijing 100084, China \\
    \textsuperscript{\rm 3} Quantum Science Center of Guangdong-Hong Kong-Macao Greater Bay Area, Shenzhen 518045, China \\
    $^{\dagger}$felixxinwang@hkust-gz.edu.cn
}

\usepackage{bibentry}
\begin{document}

\maketitle

\begin{abstract}

Balancing trainability and expressibility is a central challenge in variational quantum computing, and quantum architecture search (QAS) plays a pivotal role by automatically designing problem-specific parameterized circuits that address this trade-off. In this work, we introduce a scalable, training-free QAS framework that efficiently explores and evaluates quantum circuits through landscape fluctuation analysis. This analysis captures key characteristics of the cost function landscape, enabling accurate prediction of circuit learnability without costly training. By combining this metric with a streamlined two-level search strategy, our approach identifies high-performance, large-scale circuits with higher accuracy and fewer gates. We further demonstrate the practicality and scalability of our method, achieving significantly lower classical resource consumption compared to prior work. Notably, our framework attains robust performance on a challenging 50-qubit quantum many-body simulation, highlighting its potential for addressing complex quantum problems.
\end{abstract}
\section{Introduction}

Quantum computing~\cite{Preskill2018, Zhong2020, Bluvstein2024} has revolutionized computational science, significantly impacting quantum machine learning. 
% This paradigm leverages quantum mechanics to address problems intractable for classical systems. 
% A core concept in QML is learning quantum states, analogous to learning probability distributions in classical machine learning~\cite{Kearns1994}.  
Variational Quantum Eigensolver (VQE) has emerged as a powerful approach, leveraging parameterized quantum circuits (PQCs) to exhibit robustness against noise and adaptability to near-term noisy intermediate-scale (NISQ) devices~\cite{Bharti2022, Cerezo2021a}.  By optimizing these PQCs, VQEs enable the preparation of target quantum states, facilitating applications in condensed matter physics~\cite{Zheng2017, Qin2020, Xu2024} and combinatorial optimization~\cite{Farhi2014,Zhou2020}. 
However, the performance of VQEs is critically dependent on the architecture of the underlying PQC. This dependence necessitates the automated design of efficient quantum circuits, a crucial research area known as quantum architecture search (QAS).

Recent years have witnessed the emergence of several QAS methods, each offering unique strategies for optimizing quantum circuit design~\cite{williams1998automated, las2016genetic,romero2017quantum,lamata2018quantum,fosel2021quantum, du2020quantum, huang2022robust, ostaszewski2021reinforcement, zhang2021neural,kuo2021quantum, du2022quantum,wang2022quantumNAS, zhang2022differentiable, he2023gsqas, he2024training, patel2024curriculum}. These methodologies can be broadly categorized into several classes. Reinforcement learning (RL) based QAS methods explore circuit configurations through intelligent agents~\cite{fosel2021quantum,ostaszewski2021reinforcement,kuo2021quantum,patel2024curriculum}, while evolutionary algorithms mimic natural selection to evolve high-performing circuits~\cite{las2016genetic,romero2017quantum,lamata2018quantum,huang2022robust}. Predictor-based QAS~\cite{Zhang2021,he2023gsqas} uses machine learning to predict circuit performance, avoiding exhaustive training. However, RL-based and evolutionary algorithms typically require extensive training, whereas predictor-based QAS depends critically on representative training data and accurate ground-truth evaluations. Consequently, all these approaches face computational cost, especially for large-scale quantum circuits.

To improve the efficiency of QAS, research has increasingly focused on SuperNet-based and zero-shot QAS approaches. In the SuperNet framework, \cite{wang2022quantumNAS, du2022quantum} introduced a method that first trains a SuperCircuit and then utilizes it to estimate the performance of candidate circuits. On the other hand, inspired by zero-shot neural architecture search in classical machine learning~\cite{li2024zero}, \cite{he2024training} proposed Training-Free Quantum Architecture Search (TF-QAS), which employs zero-cost proxies such as path-based and expressibility-based metrics.
The path-based proxy quantifies circuit complexity via path count in its directed acyclic graph representation, while the expressibility-based proxy measures its ability to explore Hilbert space uniformly.
This measure allows TF-QAS to identify superior circuits while significantly reducing computational time compared to traditional QAS methods that depend on iterative training.

Despite their advantages, these methods still face limitations in scalability and applicability to large-scale quantum systems. As system size increases, TF-QAS requires expressibility evaluation, leading to significant classical memory consumption. SuperNet-based methods~\cite{wang2022quantumNAS, du2022quantum} suffer from the overhead of training supernets, resulting in high computational costs. Additionally, TF-QAS relies on simple proxies that may fail to capture the complexities of circuits needed for larger tasks, while SuperNet-based methods use fixed layouts for entanglement patterns, making it challenging to achieve both expressivity and trainability in VQEs. These two challenges leave the performance of current QAS methods on large-scale quantum circuits largely unexplored and uncertain, highlighting the need for further investigation and improvement.

The contribution of our paper is summarized as follows,
\begin{itemize}
    \item We propose a scalable zero-shot QAS framework. It first employs a layerwise search to efficiently identify the necessary entanglement pattern. Then through redundancy elimination, our method removes unnecessary gates. Both phases are guided by the relative fluctuation of landscapes, a performance predictor that ranks circuits based on their learnability. Notably, this metric can be efficiently computed via stabilizer formalism.
    \item Using the classically simulable framework, we compare several many-body systems of varying sizes and observe an average improvement of 69.8\%. We further simulate random Hamiltonians to evaluate the generalization capability of our method, demonstrating that our method achieves an average accuracy improvement of 54.3\%.
    \item Furthermore, we estimate the classical resource consumption (memory and time) in both the search and performance ranking modules. Our method achieves a 1.41x improvement in classical memory usage, due to its classical simulability, and demonstrates a 25.23x reduction in search time.
    \item We extend our analysis to large-scale VQE applications by evaluating three 50-qubit Hamiltonians, showing that the discovered circuits have an optimality gap of only 0.01 compared to the ideal solution.
\end{itemize}
% In this work, we propose a scalable zero-shot QAS framework that integrates \textbf{relative fluctuation} (RF) as a metric for evaluating quantum circuit architectures. RF measures landscape fluctuations relative to standard learnable landscapes and serves as an efficient predictor of a circuit's learnability \cite{zhang2024predicting}. By combining RF with a hybrid macro-micro search strategy, our method surpasses TF-QAS in discovering circuits with higher accuracy and fewer quantum gates while maintaining lower memory consumption. Additionally, our approach exhibits strong scalability, as RF eliminates the need for iterative training and interaction with quantum computers. Its efficient classical simulability further extends its applicability to circuits with over 50 qubits, making it a highly practical solution for large-scale quantum systems. These advancements position our method as a robust solution for scalable and efficient quantum architecture design in the NISQ era.

% \begin{figure}
%     \centering
%     \includegraphics[width=1\linewidth]{}
%     \caption{Caption}
%     \label{fig:enter-label}
% \end{figure}

\section{Preliminary}\label{sec:preliminary}

\subsection{Background}
\textbf{Quantum Computing Basic.} To manipulate information, quantum computing (QC) systems use quantum gates to modify the state vectors of qubits. Analogous to universal gates in classical computing, QC systems generally support a set of single-qubit rotation and two-qubit gates. These gates are capable of acting on one or two qubits simultaneously and have been demonstrated to express any arbitrary quantum circuit~\cite{Barenco_1995}. Common single-qubit rotation gates include $R_x(\theta)=e^{-i\theta X/2}$, $R_y(\theta)=e^{-i\theta Y/2}$, $R_z(\theta)=e^{-i\theta Z/2}$, which are in the matrix exponential form of Pauli matrices,
\begin{equation}
    X = \begin{pmatrix}
        0 & 1 \\ 1 & 0
    \end{pmatrix},\quad
    Y = \begin{pmatrix}
        0 & -i \\ i & 0 \\
    \end{pmatrix},\quad
    Z = \begin{pmatrix}
        1 & 0 \\ 0 & -1 \\
    \end{pmatrix}.
\end{equation}
Common two-qubit gates include controlled-Z gate $\text{CZ}$ and controlled-X gate $\text{CNOT}$, which can generate quantum entanglement.  $R_{xx}$, $R_{yy}$, $R_{zz}$ gate are also frequently used, where $R_{xx}(\theta) = e^{-i\theta XX /2}$ and similarly for $R_{yy}$ and $R_{ZZ}$.

\textbf{Variational Quantum Eigensolver.} In this work, we will primarily focus on the ground state preparation problem with variational quantum eigensolver (VQE) and VQE is a method that uses a hybrid quantum-classical computational approach to find eigenvalues of a Hamiltonian. For an $N$-qubit system Hamiltonian $H$, it has the decomposition of the form $H = \sum_{j}\lambda_j h_j$, where $h_j$ is a Pauli string and $\lambda_j$ is the real coefficient. We assume the input state to be $|0\rangle^{\otimes N}$ and its corresponding PQC is represented as $|\psi({\bm{\theta})}\rangle = \mathbf{U}(\theta)|\bm{0}\rangle$, where the implementation of $\mathbf{U}$ using a decomposition of the required gate set and serves as the target of QAS. Then, the VQE finds eigenvalues of a Hamiltonian by measuring and minimizing the loss function, 
\begin{equation}
\mathcal{L}(\bm{\theta})=\langle\psi(\bm{\theta})|H|{\psi(\bm{\theta})\rangle}.
\end{equation}

However, the performance of VQEs depends not only on device noise but also significantly on the structure of the PQC used. First, the design of a PQC must ensure sufficient expressivity to include the ground state within the circuit's search space, rather than simply focusing on the size of the expressive space~\cite{Holmes2021}. A PQC capable of uniformly covering the entire Hilbert space can still suffer from gradient vanishing issues, commonly referred to as barren plateaus~\cite{mcclean2018barren, Zhang2024_k}. Moreover, the design must also consider the problem of local minima~\cite{Anschuetz2022,Bittel2021,Zhang2023}, which poses another trainability challenge. Additionally, manually designing PQCs is impractical due to the exponential growth of the search space with the size of the circuit. Therefore, an efficient and trainability-aware approach is essential to enable large-scale VQA applications. 
% In the following, we provide a brief overview of the literature on the automated design of PQCs.

\textbf{Quantum Architecture Search (QAS).}
The automated process of designing quantum circuits based on the specified performance criteria is known as quantum architecture search (QAS). Developments in Neural Network Architecture Search (NAS) have been a major source of inspiration for research in the field of QAS. Similar to the exploration of neural network architectures in Neural Architecture Search (NAS), the search module in QAS iteratively evaluates various quantum circuit configurations to identify those that achieve superior performance metrics. The goal is to discover quantum circuit designs that balance enhanced learnability with manageable computational costs. The search module and the performance evaluation module make up the two main structural elements of QAS and has received the majority of attention in the design and improvement of QAS algorithms.

\begin{figure*}[t]
    \centering
    \includegraphics[width=0.95\textwidth]{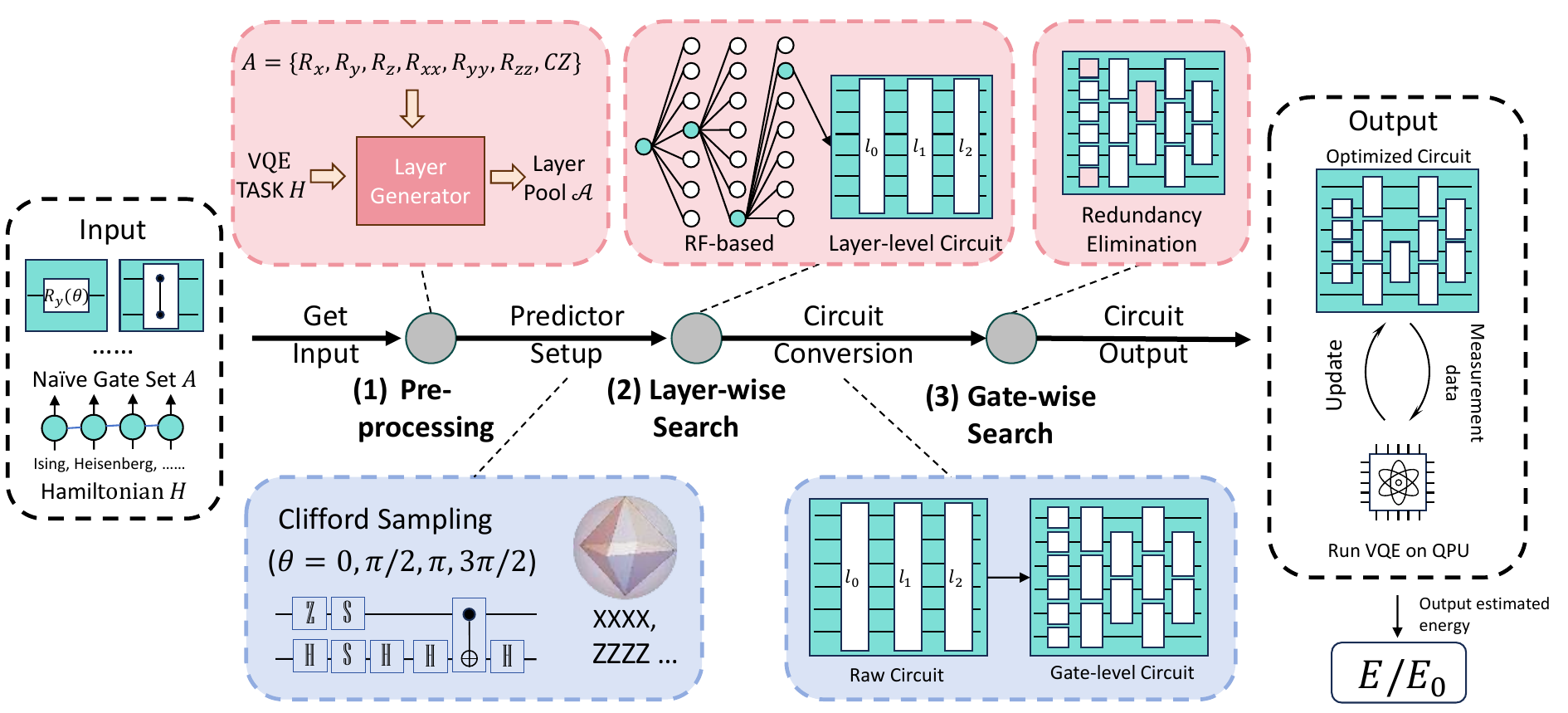}
    \caption{This figure illustrates the QAS framework. The process starts with preprocessing, using a native gate set and target Hamiltonian $H$ to search circuits. The layer-wise search stage is guided by fluctuation of landscape and it explores candidate circuits at the layer level. Then, selected layer-wise circuits undergo gate-wise search and redundancy elimination. The final optimized circuit is evaluated to calculate metrics such as the ground-state energy-to-exact ratio.}
    \label{fig:qas framework}
\end{figure*}

\subsection{Motivation}

% \paragraph{Performance of noise-resilient methods}  

\paragraph{Scalability of quantum architecture search} While existing methods offer valuable guidance for QAS in small-scale applications, scalability remains a challenge. For instance, some approaches use expressivity to evaluate circuit performance, which demands substantial classical memory to store distributions and necessitates frequent interactions with quantum devices. Additionally, SuperNet-based methods~\cite{pham2018efficient} often require a large search space, leading to significant search times even for small-scale applications. Both high classical memory consumption and excessive search time present major obstacles to achieving large-scale QAS.

% the current proxy focuses solely on expressivity, neglecting other critical factors. It is then suggested to use a unified, efficient and classical simulable metric that incorporates expressivity, barren plateaus, bad local minima, and overparameterization.

\paragraph{Large-scale application performance unexplored} Currently, quantum architecture search (QAS) for the VQE task is limited to small-scale cases, with TF-QAS restricted to 6 qubits~\cite{he2024training}, DQAS to 8 qubits~\cite{zhang2022differentiable}, and quantumNAS demonstrating scalability up to 15 qubits~\cite{wang2022quantumNAS}. While these methods perform well on the tested system sizes, their effectiveness on larger qubit systems remains unclear. Additionally, their reliance on substantial classical resources poses a significant bottleneck, making it difficult to assess their scalability. As quantum computing continues to advance, validating QAS methods on larger systems is crucial for guiding the implementation of large-scale VQE.

\section{Method}\label{sec:method}

The proposed QAS algorithm framework proposed in this paper is illustrated in Figure.\ref{fig:qas framework}. We begin by adopting a layer-wise search procedure to determine which qubit pairs require two-qubit gates. Next, gates with minimal performance impact are systematically removed during the redundancy elimination phase.

\textbf{Layer Generator.} Once the native gate set is determined, careful consideration must be given to its arrangement within the gate layers. For single-qubit gates, their relatively low computational overhead allows for the generation of transversal single-gate layers. However, the creation of layers for two-qubit gates necessitates more specialized techniques. A typical configuration for such layers follows a structure like (1,2), (3,4), and so on~\cite{he2024training}. This arrangement enhances hardware efficiency by aligning with the physical layout of many quantum devices. However, for general VQE tasks, it may not always yield optimal performance.

% The reason is that it fails to fully exploit the entanglement and parallelism that are crucial for the VQE algorithm. The VQE relies on efficient exploration of the parameter space to find the ground state of a quantum Hamiltonian, and a more entangled and parallel structure can enhance this exploration process.

To improve this, we formulate the arrangement of two-qubit gates using a graph representation. Each qubit is represented as a node in the graph, and the relationship between two nodes, \( q_1 \) and \( q_2 \), is represented by an edge with a weight. The weight of the edge is determined by the interaction between \( q_1 \) and \( q_2 \) in the VQE task. A higher weight indicates that the entanglement between these two qubits is more significant. Consequently, the generation of the layer can be transformed into a maximum cardinality matching problem. 
We use the Weighted Blossom Algorithm~\cite{edmonds1965paths} to generate the gate layer of two quadrants. 
% We modify a portion of the Blossom Algorithm~\cite{edmonds1965paths} to achieve maximum matching in a weighted graph, specifically for generating the gate layer of two quadrants.
Inspired by the brickwall layer, we perform two layers of matching using this algorithm. In the first layer, we find the maximum cardinality matching of the graph. For the second layer, we remove the edges that were matched in the first layer and apply the Blossom Algorithm again to find a new matching. With this method, our layer generator effectively leverages the characteristics of the VQE task, leading to improved performance, as demonstrated in the evaluation section.

% \CH{@HK check and revise}
\textbf{Classically Simulable Predictor.} The performance evaluation metric we use is referred to as the landscape fluctuation, defined by the standard deviation of the approximately normalized cost function:
\begin{equation}
    \sigma = \frac{\sqrt{\text{Var}_{\Xi}[\mathcal{L}]} }{ \|\pmb{\lambda}\|_1},
\end{equation}
where $\Xi$ represents the uniformly distributed ensemble of parameter points. 
% This metric captures valuable information about the landscape of the cost function. 
To account for the effect of the number of tunable parameters $M$ in PQCs, the metric incorporates a scaling factor based on the parameter count, i.e., $\sigma_0 = 1/\sqrt{2M}$, which originates from the comparison with standard learnable landscapes in variational quantum circuits. By combining these two components, the relative fluctuation is defined as:
\begin{equation}
    \tilde{\sigma} = \frac{\sigma}{\sigma_0}. 
\end{equation}
which can diagnose several key features of PQCs, including barren plateaus, bad local minima, insufficient expressibility, and overparameterization.

\begin{figure}[h]
    \centering
    \includegraphics[width=1\linewidth]{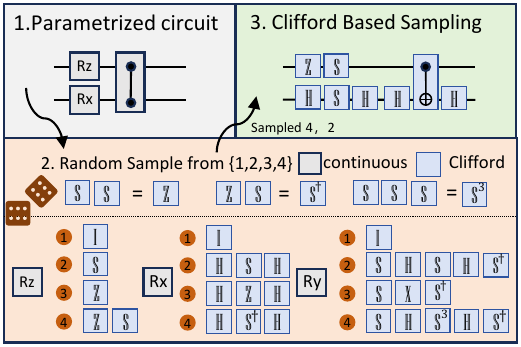}
    \caption{Procedure for estimating relative fluctuation via Clifford sampling.}
    \label{fig:clifford_sampling_process}
\end{figure}

Notably, RF only involves second-order moments, allowing the predictor to be fully classically simulated using Clifford random circuits for standard PQCs and stabilizer initial states. This means instead of sampling rotation angles from the continuous range $[0,2\pi)$, it suffices to sample from the discrete set $\{0, \pi/2,\pi, 3\pi/2\}$ to evaluate RF using the stabilizer formalism, which can be checked by using the definition of unitary $t$-designs for $U(1)$ group. Using the $R_z$ gate as an example, the continuous integral gives
\begin{equation}\label{eq:continuous}
\begin{aligned}
    \int_{0}^{4\pi} \frac{d\theta}{4\pi} P_{t,t}[R_z(\theta)] = \sum_{k=-2t}^{2t}\lambda_{k} \int_{0}^{4\pi} \frac{d\theta}{4\pi} e^{-ik\theta/2} = \lambda_0,
\end{aligned}
\end{equation}
where $P_{t,t}(\cdot)$ is a polynomial of degree at most t in the entries of matrix $(\cdot)$ and $\lambda_k$ represents the corresponding coefficients. On the other hand, the summation over a discrete arithmetic set $\mathbb{D}$ gives
\begin{equation}\label{eq:discrete}
\begin{aligned}
    \frac{1}{|\mathbb{D}|}\sum_{\theta\in \mathbb{D}} P_{t,t}[R_z(\theta)] = \frac{1}{|\mathbb{D}|} \sum_{k=-2t}^{2t}\lambda_{k} \sum_{\theta\in\mathbb{D}} e^{-ik\theta/2}.
\end{aligned}
\end{equation}
The equivalence between Eq.~\eqref{eq:continuous} and Eq.~\eqref{eq:discrete} requires that the terms with $k\neq 0$ should all vanish. Combined with the fact that the global phase does not affect the expectation value, it can show the discrete set $\{0, \pi/2, \pi, 3\pi/2\}$ suffices to compute the landscape fluctuation.
We summarize the rules for transforming a standard PQC into its Clifford-based estimator in Figure.~\ref{fig:clifford_sampling_process} for simplicity. The results can be similarly derived for commonly used two-qubit gates such as $R_{zz}$. In such a way, this approach enables the efficient estimation of the performance of large-scale PQCs.

Finally, we acknowledge the limitation of this measure that the Clifford sampling method is not applicable in scenarios with shared parameters, such as QAOA~\cite{farhi2014quantum}. However, the landscape fluctuation can still be estimated in the continuous setting, albeit at a significantly higher computational cost or with quantum resources. In the following simulation evaluation section, we will focus on the VQE task.

\textbf{Layer-wise Search Module. } In this context, we introduce a layer-wise search algorithm, with specific details outlined in Algorithm~\ref{layer-wise search algorithm}. 
The algorithm begins with an initial circuit and iteratively adds layers from a predefined pool. Each new circuit is evaluated using a scoring function $P(\mathcal{C}_a)$, which quantifies its ability to learn the ground state of the given Hamiltonian $H$. In a detail, the scoring function $P$ is defined as,
\begin{align}
    P(\mathcal{C}_a)=\tilde{\sigma}(\mathcal{C}_a) \times \alpha(\mathcal{C}_a),
\end{align}
where $\mathcal{C}_a$ is quantum circuit with newly added layer $a$, $\tilde{\sigma}$ is relative fluctuation and $\alpha$ is a decay term.
The algorithm then selects the best performing layer configuration at each step and updates the current circuit accordingly.
Since the enhancement of a circuit's learning capabilities through the mere repetition of identical circuit layers is inherently limited, we introduce a decay factor into the design. The decay term is defined as $\alpha(\mathcal{C}_a) =\delta^{d_a}$, where $d_a$ represents the number of layers within the final $k$ layers of $\mathcal{C}$ where the circuit layer $a$ occurs. In our later experiments, we set $\delta=0.8$ and $k=5$.
A stopping criterion is included to terminate the process early if the circuit's performance exceeds a threshold $ 1-\epsilon$ and meets the minimum layer requirement $ l_{min} $. This balances accuracy and complexity, ensuring an efficient and effective quantum circuit design. The complexity of layer-wise searching is $O(poly(N)|\mathcal{A}|l_{max})$, where $poly(N)$ is the complexity of Clifford sampling with respect to the number of qubit $N$ and $|\mathcal{A}|$ represents the number of layer candidate.

\begin{algorithm}[h]
    \caption{Layer-wise Searching}
    \begin{algorithmic}
    \REQUIRE Hamiltonian $H$, qubit number $n$, circuit layer pool $\mathcal{A}$, min layer number $l_{min}$, max layer number $l_{max}$
    \STATE Initial current circuit $\mathcal{C}$
    \WHILE{$l(\mathcal{C}) < l_{max}$}
    \FOR{ $a \in \mathcal{A}$}
    \STATE Generate new circuit $\mathcal{C}_a = \mathcal{C}+a$
    \STATE Compute new circuit's score $P(\mathcal{C}_a)$
    \ENDFOR
    \STATE Select the $\mathcal{C}_a$ with highest score and update $\mathcal{C}=\mathcal{C}_a$
    \IF{$P(\mathcal{C})> 1-\epsilon$ and $l(\mathcal{C})\geq l_{min}$}
    \STATE Break
    \ENDIF
    \ENDWHILE
    \RETURN $\mathcal{C}$
    \end{algorithmic}    
    \label{layer-wise search algorithm}
\end{algorithm}

We note that the high-level concept of layer-wise searching has also been explored in previous works~\cite{zhang2022differentiable,he2024training}, though without incorporating a penalty term. These studies also report that gate-wise searching frameworks, which construct quantum circuits by adding gates one at a time, tend to exhibit ``chaotic" behavior and often suffer from the barren plateau problem. Another key reason for using layer-wise searching is to reduce the computational overhead. Compared to gate-wise searching, layer-wise searching is more efficient because it involves fewer choices for gate placement and configuration, making it less computationally expensive. In the results section, we will demonstrate that this straightforward framework not only addresses these issues but also outperforms prior methods.

\textbf{Redundant Gate Elimination.}
The layer-wise search method facilitates the rapid identification of a circuit architecture that achieves satisfactory performance. However, this approach is still have its limitations. Specifically, within the circuit layers, gates are systematically placed at various predetermined locations, which may not always lead to improvements in training efficiency. The inclusion of such redundant gates can hinder optimal performance and increase the depth of the circuit. Reducing these redundant gates could significantly reduce the number of gates in the circuit and lower the quantum resources required for parameter tuning.

\begin{figure}[h]
    \centering
    \includegraphics[width=1\linewidth]{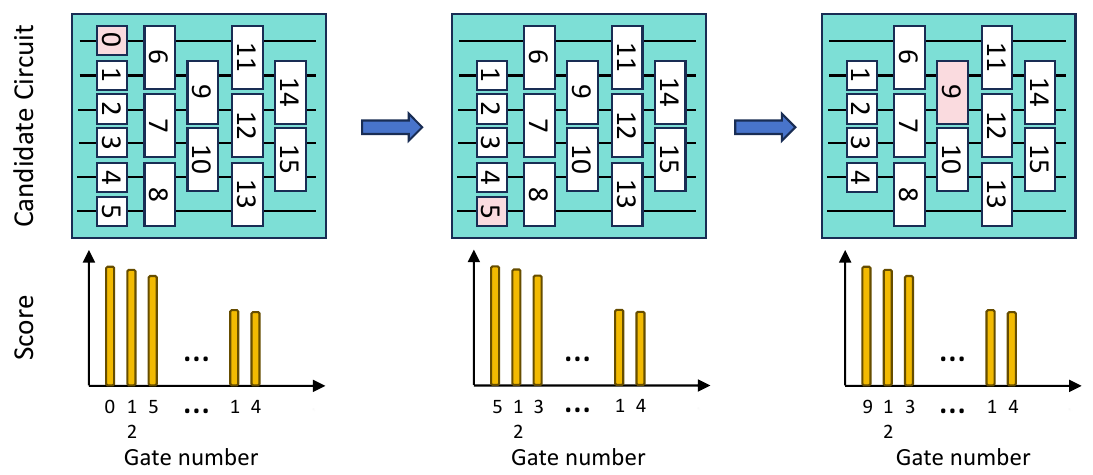}
    \caption{An illustration of the redundant gate elimination phase. The process iteratively removes target gates, evaluating the circuit's performance after each removal. The gate whose removal results in the highest score improvement is selected for elimination (marked pink in the circuit).}
    \label{fig:gate_elimination}
\end{figure}

To address this issue, we have modified Algorithm.~\ref{layer-wise search algorithm} to include a second-level search module specifically designed for eliminating redundancy gates. The illustration of this process is provided in Figure~\ref{fig:gate_elimination}.
This algorithm takes the output of the layer-wise search as its input and employs a targeted search mechanism to identify and remove redundant gates within the circuit. For each iteration, The gate whose removal results in the highest score improvement is selected for elimination. By refining the circuit architecture in this manner, the algorithm aims to reduce the redundant gate and improve the overall efficacy of the quantum circuit. The complexity of redundancy gate elimination is $O(poly(N)|\mathcal{C}| k)$, where $poly(N)$ represents the cost of Clifford sampling with respect to the number of qubit $N$, $|\mathcal{C}|$ denotes the number of gates in circuit $\mathcal{C}$ and $k$ denotes the number of iterations, representing the number of times the process of removing predefined gates is executed.

% \WU{The score function $P'$}

% \begin{algorithm}[h]
%     \caption{Redundant Gate Elimination}
%     \begin{algorithmic}
%     \REQUIRE Hamiltonian $H$, circuit $\mathcal{C}$, max removal number $r_{max}$, removal number per step $r_{step}$
%     \STATE Initial removal counter $r_{cnt}$
%     \WHILE{$r_{cnt}<r_{max}$}
%     \STATE Initial set $S=\{\}$
%     \FOR{ $g \in \mathcal{C}$}
%     \STATE Generate new circuit $\mathcal{C}_g = \mathcal{C}-a$
%     \STATE Compute new circuit's score $P'(\mathcal{C}_g)$
%     \STATE $S=S\cup g$
%     \ENDFOR
%     \STATE Sort the new circuits in descending order of score.
%     \FOR{$i$ in range($r_{step}$)}
%     \STATE Remove gate $S[i]$ from $\mathcal{C}$ 
%     \STATE $r_{cnt}=r_{cnt}+1$
%     \IF{$r_{cnt}>=r_{max}$}
%     \STATE Break
%     \ENDIF
%     \ENDFOR
%     \ENDWHILE
%     \RETURN $\mathcal{C}$
%     \end{algorithmic}    
%     \label{gate-wise search algorithm}
% \end{algorithm}
% \begin{algorithm}[h]
%     \caption{Redundant Gate Elimination}
%     \begin{algorithmic}
%     \REQUIRE Raw circuit $\mathcal{C}$, maximum gate removal count $M$
%     \FOR{$i$ in range(M)}
%     \FOR{$g \in \mathcal{C}$}
%     \STATE Generate new circuit $\mathcal{C}_g=\mathcal{C}-g$
%     \STATE Compute the score $P'(\mathcal{C}_g)$
%     \ENDFOR
%     \STATE Select the $\mathcal{C}_g$ with highest score    
%     \IF{$P'(\mathcal{C}_g) >\epsilon$}
%     \STATE Break
%     \ENDIF
%     \STATE Update $\mathcal{C}= \mathcal{C}_g$
%     \ENDFOR
%     \RETURN $\mathcal{C}$
%     \end{algorithmic}    
%     \label{gate-wise search algorithm}
% \end{algorithm}

\section{Experiment Setup}\label{sec:setup}
% We first introduce the general setup and the used benchmark of our simulations.

\textbf{Experiment Setting.} We evaluate the performance of our searched circuit relative to the true value $E_0$. For each searched circuit, we initialize the parameters uniformly from $[0,2\pi)$ and use the Adam optimizer with a learning rate of 0.1. The training error is averaged over 100 independent learning instances to provide statistical results. We report the searched value $E$ and assess its accuracy by calculating the ratio $E/E_0$. For each experiment, we allocate a maximum execution time of 3000s. If the program fails to complete within this limit, it is marked as out of time (O.O.T.).

\textbf{Benchmark Application Selection.} We evaluate the performance of our proposed method across different VQE tasks for finding the ground states in many-body quantum systems. We specifically consider the task of finding ground state for three different Hamiltonians. (1). \textit{1D transverse-field cluster model (Cluster)} which is defined as, $H_{Cluster}=-\sum_{j} Z_{j-1}X_{j}Z_{j+1},$ where the boundary terms are $X_0 Z_1$ and $Z_{n-1}X_n$. (2). \textit{Heisenberg Hamiltonian} with an external field, which is defined as, $H_{Heis} = -\sum_j(X_jX_{j+1}+Y_jY_{j+1}+Z_jZ_{j+1})-\sum_{j}^{}Z_j$. (3). \textit{1D transverse-field ising model (Ising)}, which is defined as,
$H_{Ising} = -\sum_jZ_jZ_{j+1}-\sum_{j}^{}X_j$. In addition, we also extend our analysis to include the random Hamiltionians.

\textbf{Native Gate Set Selection.} Following previous work in constructing the search space, we choose the native gate set as `\texttt{RXYZ2XYZ}': $A=\{R_x, R_y, R_z, R_{xx}, R_{yy}, R_{zz}, CZ\}$. Additionally, we conduct extra simulations to compare the effects of different native gate sets, as shown in Figure.~\ref{fig:native_gate_analysis}. 

\textbf{Baseline Algorithms.} We benchmark our results against different search frameworks: (1). DQAS~\cite{zhang2022differentiable}, which is based on the DARTS framework~\cite{liu2018darts}; (2). TF-QAS~\cite{he2024training}, which utilizes random circuits; (3). quantumNAS~\cite{wang2022quantumNAS}, which builds upon Supernet~\cite{pham2018efficient}; (4). In general, optimizing circuit design to achieve ideal entanglement, minimize gates, and ensure expressibility remains challenging. To illustrate energy gaps in native gate decomposition and expressibility differences, we use a Cartan-decomposition layer~\cite{khaneja2000cartan}, a block-wise universal two-qubit gate layer, referred to as 'Cartan-2'.

\textbf{Algorithm Implementation Software.} All simulations are performed using the open-source libraries TensorCircuit~\cite{zhang2023tensorcircuit}, PyClifford~\cite{Hu2023}, and torchQuantum~\cite{wang2022quantumNAS} in Python. The random Hamiltonion is generated via Pennylane~\cite{bergholm2022pennylane}.

\textbf{Simulation Device.} All experiment are performed on a Windows 11 system with AMD Ryzen 5 6600H CPU and 16-GB physical memory.

\section{Evaluation Results}\label{sec:evaluation}
In this part, we evaluate the performance of our method on several variational quantum eigensolver tasks for finding the ground states of the given Hamiltonian. 

\subsection{Results in small-scale systems}
\begin{figure*}[ht]
    \centering
    \includegraphics[width=1\linewidth]{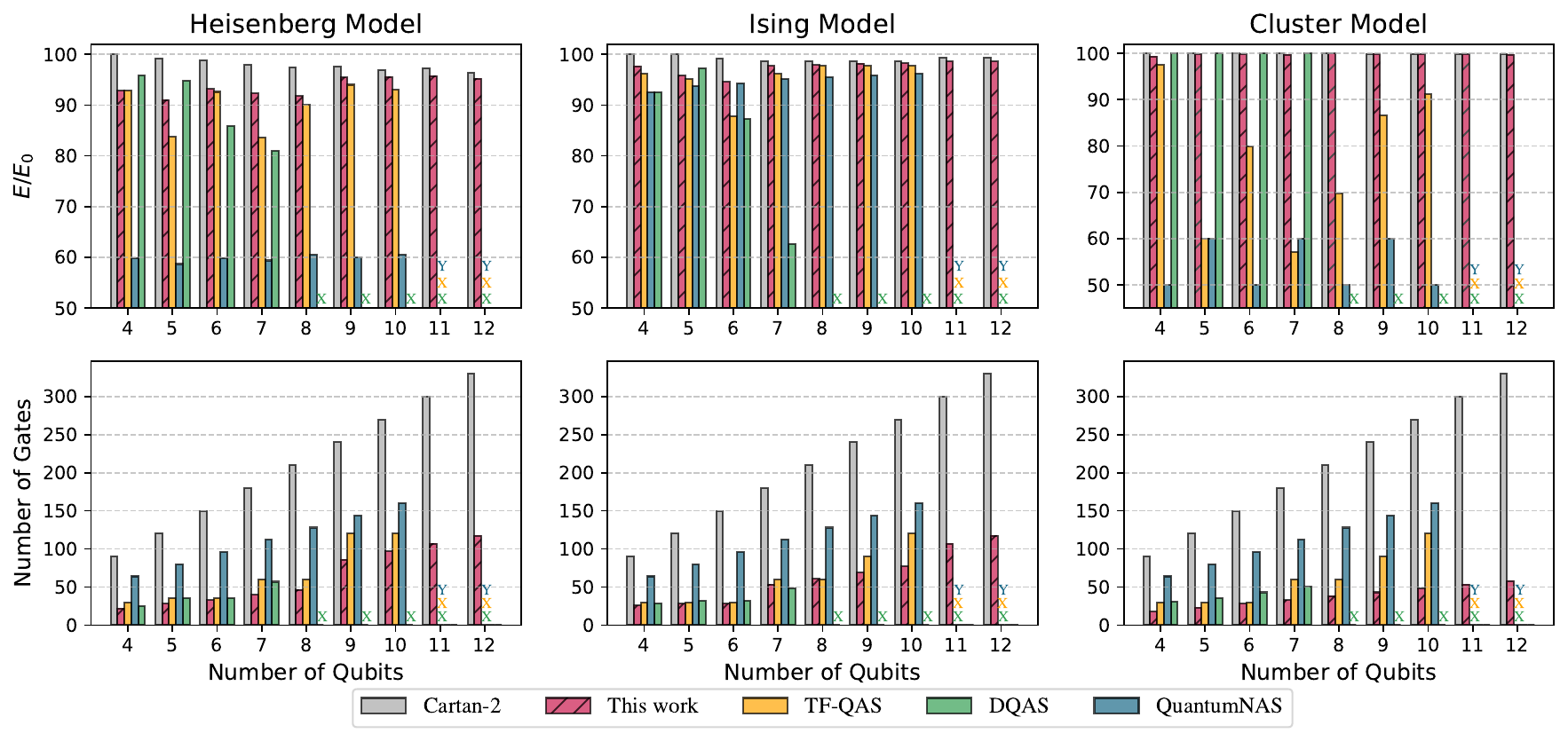}
    \caption{ Performance comparison on the three benchmark Hamiltonians with varying numbers of qubits (higher is better). The top plot shows the closeness to the ground state energy ($E/E_0$). "O.O.M." indicates cases where the classical memory requirements exceeded the computational limits. The bottom plot illustrates the number of gates used (lower is better).}
    \label{fig:vary_qubits}
\end{figure*}
\paragraph{Fixed Qubit Count.} The first example we consider is all three benchmarks with the fixed number of qubit. The Hamiltonian we chosen is Cluster, Heisenberg model and Ising. All examples are considering with the number of qubit as $N=6$. The results are shown in Table \ref{layer-wise results}, where we represent the number of gates (\#gates) after running QAS algorithms and the closeness to the ground truth ($E/E_0$).

\begin{table}[h]
    \centering
    \caption{Training result of different Hamiltonians ($N=6$).
    }
    \begin{tabular}{c|c|c|c}
    \hline
         VQE task& Method &\#gates & $E/E_0$  \\
         \hline
         \multirow{4}{*}{Cluster} & This work & \textbf{11} & \textbf{100.0\%} \\
         & TF-QAS & 30 & 80.0\% \\
         & DQAS & 28 & 100.0\% \\
         & quantumNAS &96 & 60\% \\
         & Cartan-2 & 150 & 100.0\% \\
         \hline
         \multirow{4}{*}{Heisenberg} &  This work & \textbf{33} & \textbf{93.2\%} \\
         & TF-QAS & 36 & 92.6\% \\
         & DQAS & 36 &  85.9\% \\
         & quantumNAS &96 & 69\% \\
         & Cartan-2 & 150 & 98.8\% \\
         \hline
         \multirow{4}{*}{Ising} &  This work &  \textbf{18} & \textbf{94.4\%}  \\
         & TF-QAS & 36 &  88.3\% \\
         & DQAS & 28 & 82.3\% \\
         & quantumNAS &96 & 94.3\% \\
         & Cartan-2 & 150 & 98.9\% \\
        \hline
    \end{tabular}
    \label{layer-wise results}
\end{table}

When comparing our method to all baseline models, our method consistently achieves high accuracy with a significantly smaller number of gates. Notably, our method achieves a competitive performance with fewer gates compared to the undecomposed method Cartan-2, as seen in its 100\% accuracy for Cluster with only 11 gates. For instance, our method achieves 100.0\% accuracy with only 11 gates in Cluster, whereas TF-QAS and DQAS require 30 and 28 gates, respectively, to achieve 98.0\% and 100.0\% accuracy. This observation continues across other scenarios. For the Heisenberg model, our method achieves 93.2\% accuracy with 33 gates, while TF-QAS requires 36 gates to achieve 92.6\%, and DQAS achieves 85.9\% with the same number of gates. 
Unlike the unstable performance of quantumNAS, our method demonstrates greater stability.
Although the performance of our method is slightly lower than that of Cartan-2, it closely approaches the accuracy achieved by Cartan-2 across all tasks. This similar performance with significantly lower resource requirements demonstrates the effectiveness of our method in approximating more complex and resource-intensive approaches like Cartan-2.

\paragraph{Varying Qubit Count.} In the second case, we evaluate performance across system sizes for a fixed Hamiltonian (Figure~\ref{fig:vary_qubits}). The top plot shows closeness to the ground-state energy, where higher values indicate better performance, while the bottom plot compares gate counts across methods.

Our method consistently delivers competitive accuracy across all qubit numbers while utilizing fewer gates than other methods. For example, our method achieves 93.2\% accuracy for 6 qubits, close to Cartan-2's superior 98.8\%, despite Cartan-2 requiring significantly more gates. DQAS exhibits relatively lower accuracy, particularly for larger qubit numbers, and encounters ``O.O.M." (Out of Memory) failures beyond 8 qubits due to its high computational demands. In contrast, our method showcases a distinct trend in circuit search performance as the size of quantum systems increases, maintaining stable accuracy without the decay observed in DQAS as the system size grows. Overall, our method achieves an average improvement of $69.8\%$ across all tasks.

We also note that for smaller cases our method and TF-QAS do not outperform Cartan-2. However, as the system size increases, our method demonstrates superior scalability and efficiency, maintaining competitive performance with significantly fewer gates compared to Cartan-2, which highlights the strength of our approach in larger-scale quantum applications. For TF-QAS, their results exhibit instability across different cases due to its reliance on a sampling-based method. In contrast, our method consistently outperforms TF-QAS's best-selected circuit while requiring fewer gates.

\begin{figure*}
    \centering
    \includegraphics[width=1\linewidth]{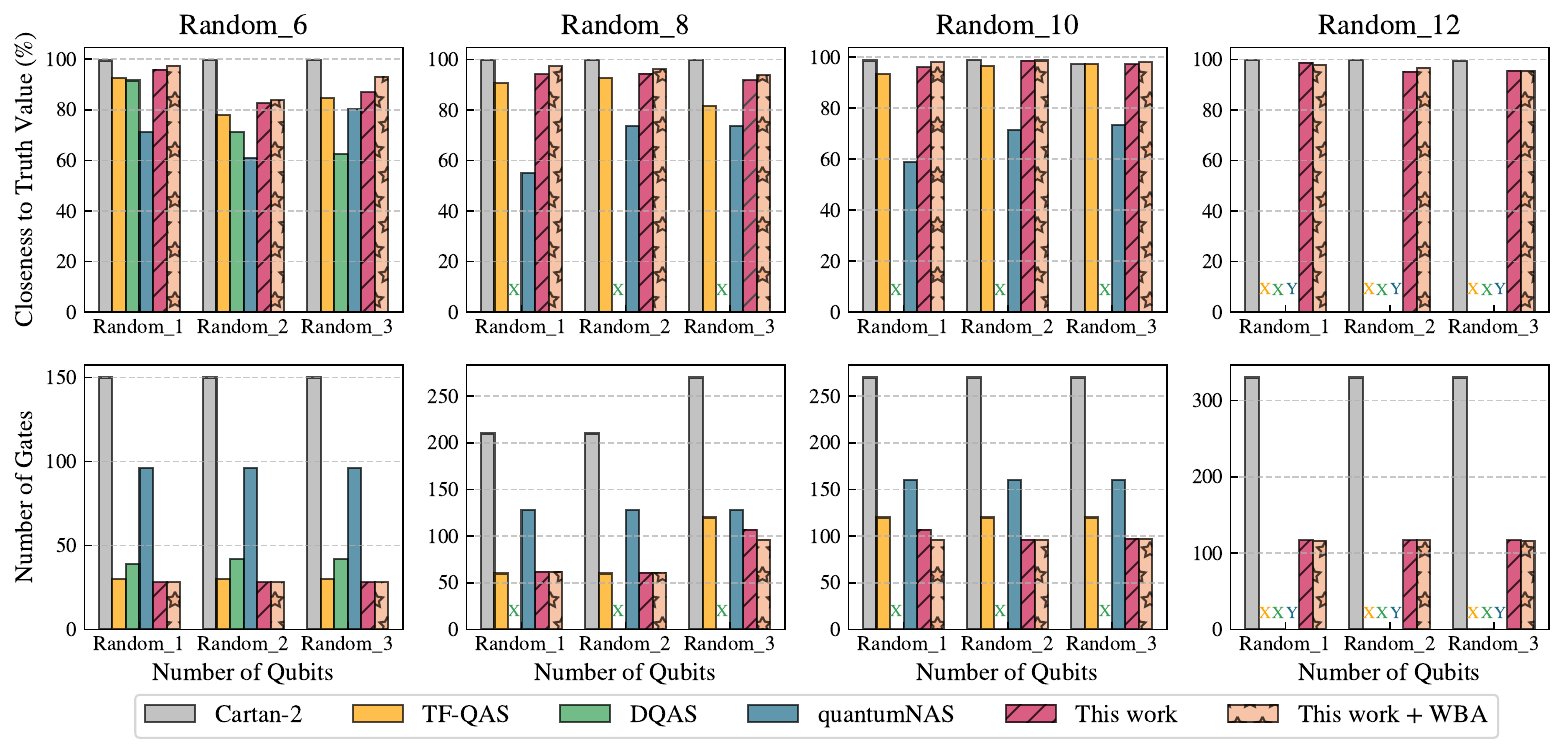}
    \caption{Performance on Random Hamiltonians (higher is better) and a breakdown of our proposed algorithm's performance. Our method achieves higher accuracy while maintaining a lower gate count and it has a more stable performance on random Hamiltonian. We also show the breakdown performance of our layer generator based on weighted blossom algorithm.}
    \label{fig:random_ham}
\end{figure*}

\textbf{Generalization and Local Minima Analysis.} Beyond specific Hamiltonians, we also assess our method on randomly generated ones to evaluate its generalizability. In detail, we consider the following \textit{locally randomized scrambled} Hamiltonian $H_{r} = V^\dagger H_Z V$, where $H_Z = -\sum_j Z_j$ and $V$ is a reversed random 1D brickwall circuit of depth $D$. A similar Hamiltonian was considered in~\cite{Anschuetz2022, zhang2024predicting} to analyze the local minima phenomenon, as its expressibility can be easily evaluated to ensure that the ansatz covers the ground state. This type of Hamiltonian can also aid in investigating local minima awareness in the QAS process. We then randomly generate locally randomized scrambled Hamiltonian with $D=1$ and test the performance our algorithm. These Hamiltonians typically contain around 500 Pauli strings for $N=6$ and $N=8$, and around 1,000 Pauli strings for 
$N=10$ and $N=12$.

The results are shown in Figure~\ref{fig:random_ham}.
Our method consistently demonstrates superior performance compared to three baselines across all tasks. For all cases, our method achieves an better accuracy on these Hamiltonian. For instance, in the case of $N=6$, our method achieves 95.9\% accuracy with 28 gates for Random\_1, surpassing TF-QAS 92.7\% and DQAS 91.6\%, both of which require more gates. Similarly, for Random\_2, our method attains 82.6\% accuracy, outperforming TF-QAS 77.9\% and DQAS 71.2\%, while maintaining the lowest gate count. This observation remains consistent across random Hamiltonians with varying qubit counts. 

Unlike the previous results on the Ising model, quantumNAS struggles to optimize the ansatz for these random Hamiltonians due to its restricted layer generation, which relies on a single layout. Consequently, its accuracy remains limited to approximately 60\% to 70\%.  Across all cases, our method achieves an average accuracy improvement of 54.3\%. For the Cartan-2 method, its superior performance can be attributed to its high expressivity, which ensures that the target ground state is contained within its parameterized circuit structure. Additionally, the random Hamiltonian is generated in a reversed manner, further aligning with the expressivity capabilities of Cartan-2. Although the gate reduction into a native gate set diminishes its expressivity, making it unable to fully recover its original capacity, our method still demonstrates better performance than other methods. This highlights the robustness of our method in capturing the essential features of the target state, even under such constraints.

\textbf{Breakdown Performance. } Figure~\ref{fig:random_ham} also presents the detailed performance breakdown of our techniques proposed in Section~\ref{sec:method}, including an ablation study on the use of the weighted blossom algorithm. We compare its effectiveness against the case where it is not applied, demonstrating that incorporating the weighted blossom algorithm enhances the identification of key correlations between Hamiltonian terms. This leads to an improvement of 25.5\% in achieving the true value or ground-state energy.

\begin{figure}[h]
    \centering
    \includegraphics[width=1\linewidth]{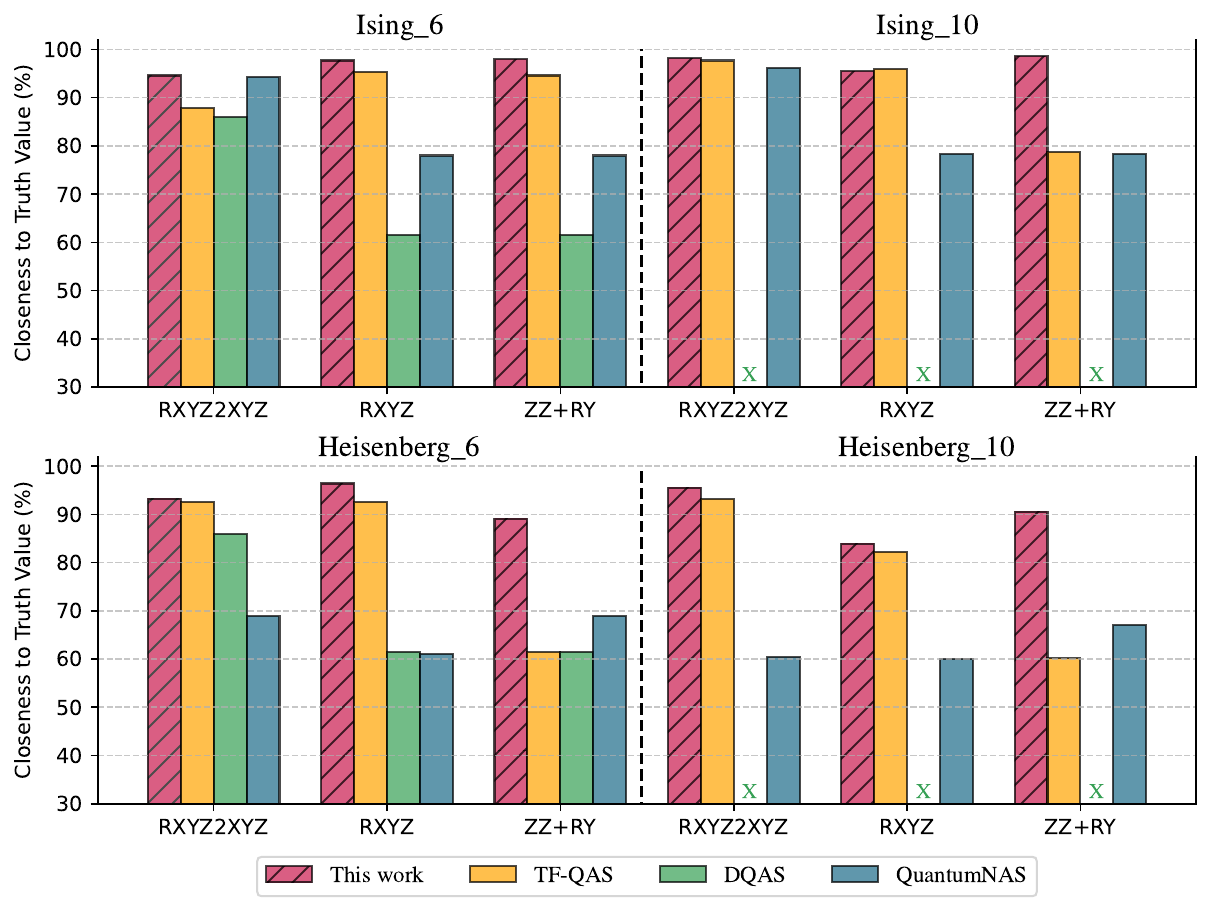}
    \caption{Native Gate Set Analysis. Our method demonstrates robustness across three different gate sets. }
    \label{fig:native_gate_analysis}
\end{figure}

\textbf{Native Gate Set Analysis.} We also conduct simulations to evaluate the effect of different native gate sets. In addition to the previously considered set `\texttt{RXYZ2XYZ}', we explore alternative sets that have demonstrated strong performance in quantumNAS~\cite{wang2022quantumNAS}: `\texttt{RXYZ}' ($A_1=\{ Rx, Ry, Rz, CZ\}$) and `\texttt{ZZ+RY}' ($A_2 = \{R_{zz}, Ry\}$). We select two example Hamiltonians with different system sizes ($N=6$ and $N=10$) to evaluate the performance of various gate sets. The results are shown in Figure~\ref{fig:native_gate_analysis}. Overall, our method demonstrates robustness across different native gate sets, effectively adapting to various gate configurations while maintaining strong expressibility and achieving higher accuracy compared to the three baselines. Compared to TF-QAS, our method achieves an average improvement of $1.09$x, and it outperforms DQAS by an average of $1.165$x. Additionally, our method surpasses the SuperNet-based quantumNAS method by an average of $1.272$x, exhibiting greater stability across different settings.

\begin{figure}[h]
    \centering
    \includegraphics[width=0.9\linewidth]{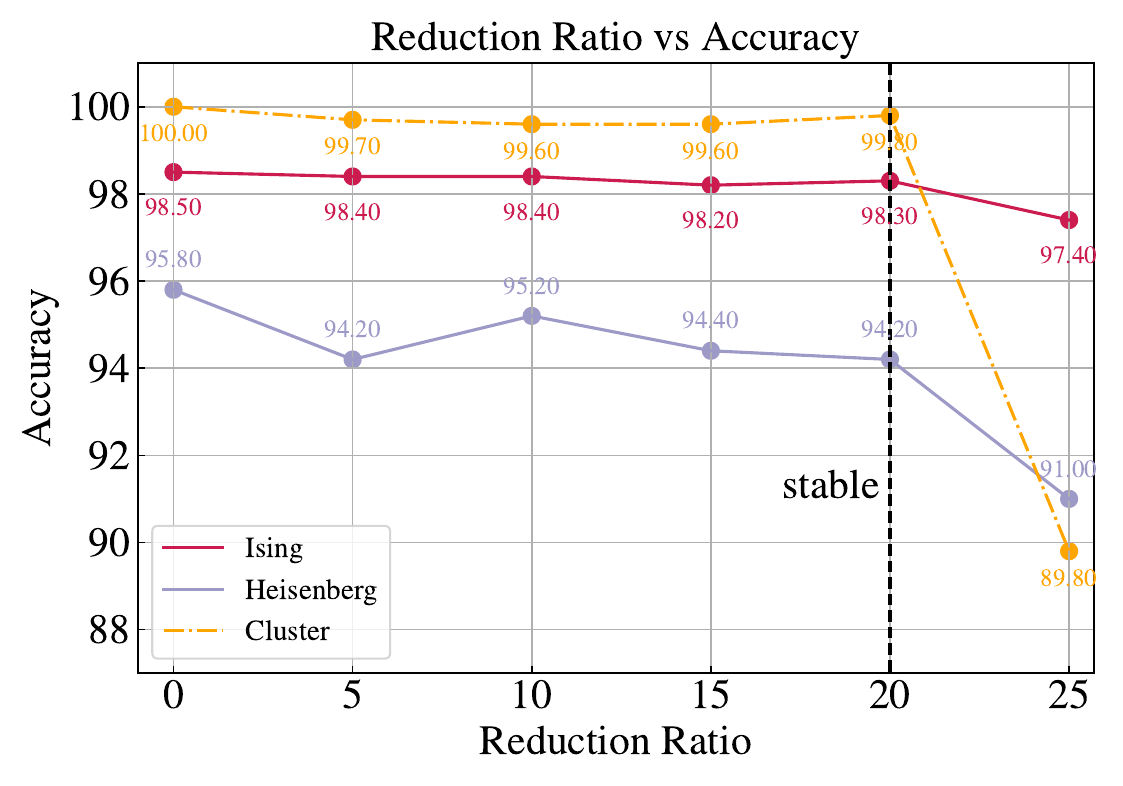}
    \caption{Circuit performance across gate reduction ratios. Accuracy remains stable below 20\% but drops sharply beyond this threshold.}
    \label{fig:redundancy_elimination}
\end{figure}

\textbf{Redundancy Elimination Analysis.}
Furthermore, we analyze the effect of redundant gate elimination for three Hamiltonians with a qubit size of 10. We analyze redundancy elimination across gate reduction ratios ranging from 0\% to 25\%. The results are presented in Figure~\ref{fig:redundancy_elimination}. Overall, the accuracy remains largely unchanged within an elimination ratio of $0\%$ to $20\%$. However, beyond this critical point, accuracy declines as the removal of necessary gates reduces expressibility. This simulation indicates that our method effectively identifies redundancy in layer-wise search, with a 20\% reduction ratio achieving an optimal balance.

\textbf{Hardware-aware Analysis.} We further conduct analysis on the topology of superconducting devices, assessing compiled gate counts and circuit depth. For compilation, we use IBM’s 64-qubit Manhattan processor~\cite{Mooney_2021} with a heavy-hex coupling topology, along with earlier IBM processors  Melbourne~\cite{kusyk2021survey} and Montreal~\cite{yu2023simulating}. We use Heisenberg and Ising model with 10 qubit as an example. quantumNAS incurs higher gate counts due to its fixed ansatz template, with the main overhead arising from gate decomposition and additional SWAP operations. In contrast, our method's layer generator effectively identifies key transformation costs, optimizing circuits for superconducting hardware and resulting in lower gate counts and reduced circuit depth.

\begin{figure}[t]
    \centering
    \includegraphics[width=1\linewidth]{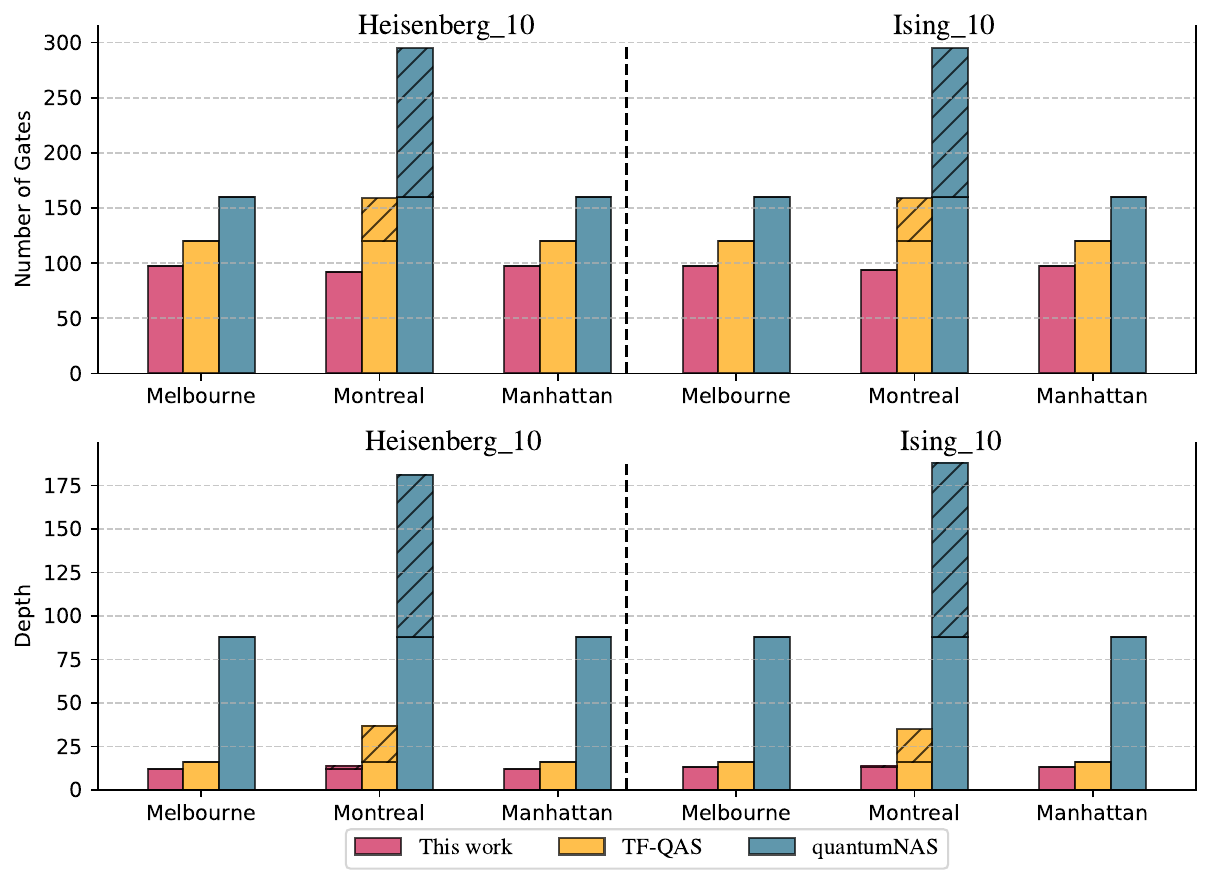}
    \caption{Superconducting hardware-aware Analysis.}
    \label{fig:topology_analysis}
\end{figure}

\subsection{Generalization to larger systems}

Next, we will extend our investigation of the QAS method to larger systems, addressing aspects that have received limited attention in previous studies. 

\begin{figure}[ht]
    \centering
    \includegraphics[width=0.9\linewidth]{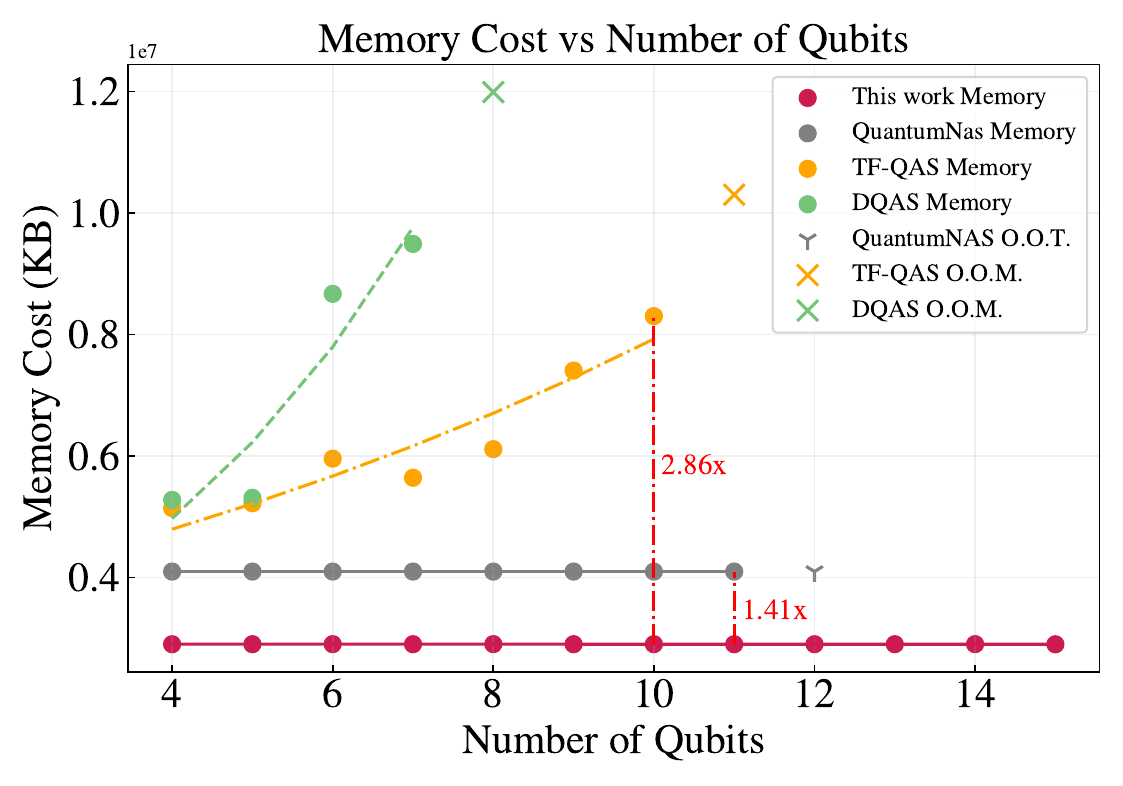}
    \caption{ Memory cost comparison (lower is better). Solid points indicate actual memory usage, while dashed lines represent fitting trends. `Out of Memory' cases are marked with X, and `Out of Time' cases are marked with Y. }
    \label{fig:memory_cost}
\end{figure}

\textbf{Classical Memory Consumption.} One of the critical challenges in performing QAS on larger systems is the significant memory cost associated with searching over the native gate space and predicting the performance of the generated circuits. We then estimate the memory consumption for each method. The memory usage during the QAS process is illustrated in Figure~\ref{fig:memory_cost}.

Our analysis reveals that the memory demand of our algorithm remains stable as qubit count $n$ increases, whereas DQAS and TF-QAS experience significant growth. DQAS scales steeply, exceeding device limits at $n=8$, while TF-QAS shows variability due to its random circuit generation and selection process. The most resource-intensive subroutine of the TF-QAS method is computing expressibility, which requires storing the entire state distribution and comparing it to the Haar distribution using the Kullback-Leibler divergence. Our analysis shows that TF-QAS is expected to exceed the memory capacity of our device when $n=11$. Compared to TF-QAS, our approach reduces memory consumption by 2.86×. Similarly, while quantumNAS maintains constant memory usage, our method still achieves a 1.41× reduction.

\begin{figure}[ht]
    \centering
    \includegraphics[width=0.9\linewidth]{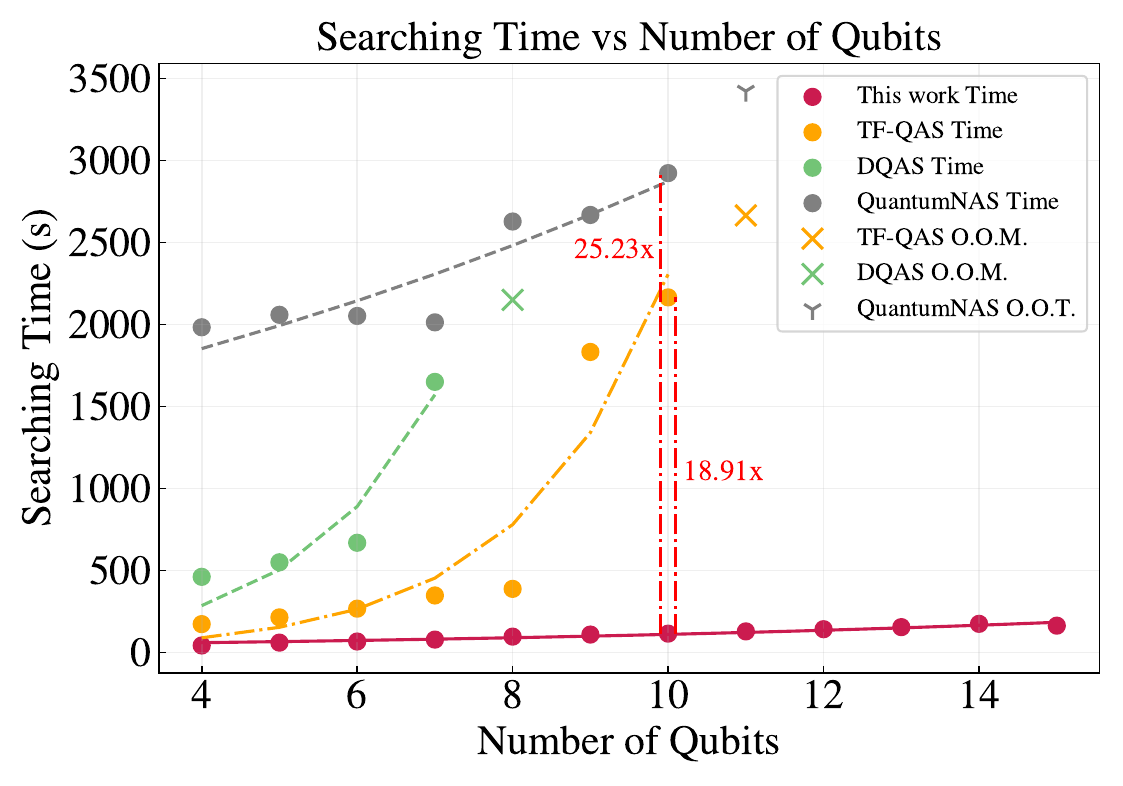}
    \caption{Searching and evaluation time comparison (lower is better). }
    \label{fig:time_cost}
\end{figure}

\textbf{Ansatz Searching and Ranking Time.}
Furthermore, we analyze an additional aspect of classical resource consumption by estimating the computational time required for searching and performance ranking. This evaluation is also critical for scaling the QAS process to large-scale applications. Using the Heisenberg Hamiltonian as an example, we evaluate systems with qubit numbers ranging from 4 to 15. The results are shown in Figure~\ref{fig:time_cost}.

Similar to classical memory consumption, the search time for TF-QAS and DQAS increases significantly with system size, whereas quantumNAS exhibits a slower increase but incurs a higher overall search time due to the Supercircuit bottleneck. our method combines the advantages of both approaches, demonstrating a slower search time growth with respect to system size while maintaining lower overall search time. Notably, our method achieves a maximum reduction of 25.23× compared to quantumNAS and 18.91× relative to TF-QAS.

\begin{figure}[h]
    \centering
    \includegraphics[width=0.8\linewidth]{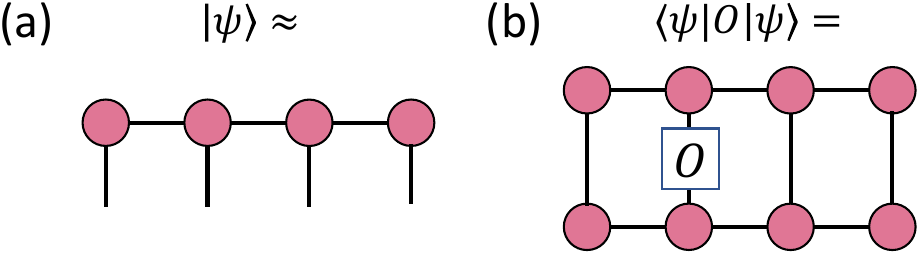}
    \caption{(a) Matrix Product State. (b) Computation of the expectation value for local observable $O$.}
    \label{fig:mps_example}
\end{figure}

\textbf{Performance on 50-qubit systems.} In addition to analyzing the memory cost of the search and evaluation modules, it is also crucial to evaluate the accuracy of our method in large-scale systems. Since the ground truth values cannot be efficiently obtained through brute-force matrix-vector representation, we leverage matrix product states (MPS) as a reliable computational framework~\cite{Cirac_2021}. We illustrate the basic idea in Figure~\ref{fig:mps_example}, where the expectation value of the given Hamiltonian is expressed as the sum of observables, as shown in Figure~\ref{fig:mps_example} (b). 
We present the results for both models in Table~\ref{50qubit_example} and we exclude comparisons with TF-QAS and DQAS as both methods exceed memory limitations during searching and circuit performance ranking module.

Table~\ref{50qubit_example} summarizes the performance of our method and Cartan-2 on 50-qubit systems for the Ising and Heisenberg models. The results indicate that our method achieves an impressive performance, with an 
$E/E_0$ ratio of 97.5\% for Ising and 97.7\% for the Heisenberg model, demonstrating its effectiveness in larger qubit systems. Although Cartan-2 achieves slightly better accuracy (99.0\% for Ising and 99.2\% for Heisenberg), it requires significantly more gates compared to our method. This result shows the scalability of our method are still able to offer reliable results in large-scale quantum systems.

\begin{table}[h]
    \centering
        \caption{Performance on 50-qubit systems.
    }
    \begin{tabular}{c|c|c|c|c}
    \hline
         VQE task& Method &\#gates & $E/E_0$ & \#iter\\
         \hline
          \multirow{2}{*}{\shortstack{Ising \\ (N=50)}}
          & This work &  397 & 97.5\% &91.5\\
          & Cartan-2 & 1470 & 99.0\% & 61.8 \\
          \hline
          \multirow{2}{*}{\shortstack{Heisenberg \\ (N=50)}}
          & This work &  497 & 97.7\% &121.25\\
          & Cartan-2 & 1470 & 99.2\% & 78.65 \\
          \hline
          \multirow{2}{*}{\shortstack{Cluster \\ (N=50)}}
          & This work & 248  &99.8\% &84.51 \\
          & Cartan-2 & 1470 & 99.7\%& 75.52  \\
          \hline
    \end{tabular}
    \label{50qubit_example}
\end{table}

\begin{figure}[h]
    \centering
    \includegraphics[width=1\linewidth]{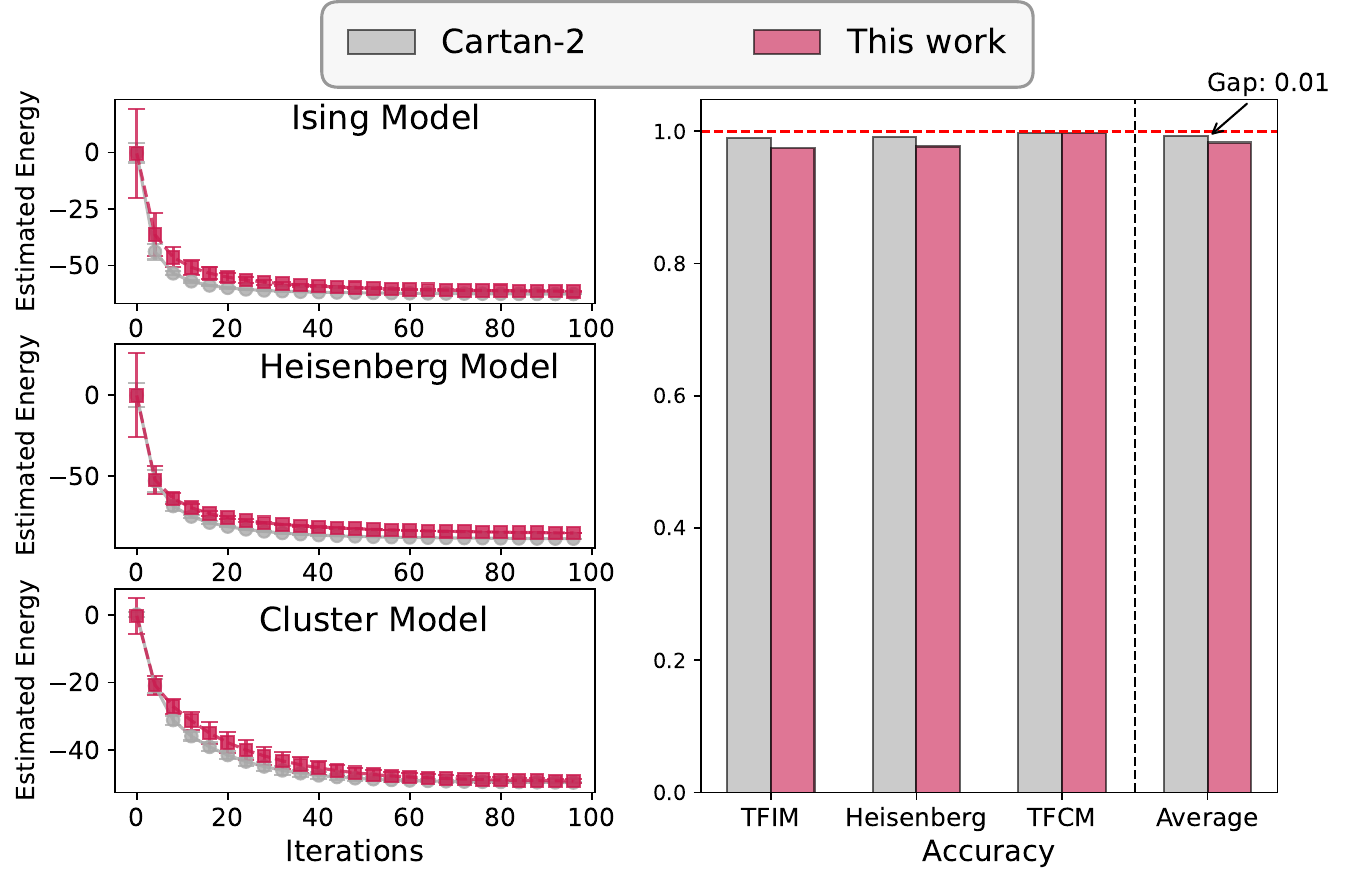}
    \caption{Training curves and final accuracy for 50-qubit cases, with error bars showing variability. Our method captures key aspects of Cartan-2 in native gate decomposition, achieving comparable performance.}
    \label{fig:training_curve}
\end{figure}

\textbf{Optimality Analysis.}
We also provide the corresponding training curves and the closeness to the truth value in Figure~\ref{fig:training_curve}. The Cartan-2 ansatz consists of many universal 2-qubit gates, each capable of simulating any arbitrary two-qubit unitary operator (up to a global phase), offering greater expressibility. However, it requires a large number of gates and needs to be decomposed into simpler, native gates. When using a given set of native gates, our method achieves a 74.1\% reduction in gate count while maintaining an approximate 1\% performance gap. These results underscore the scalability and efficiency of our method in handling large-scale quantum systems, even when using a native gate set.

\section{Related Work}\label{ref:related_work}
\textbf{Searching Module in QAS.}
\textit{Reinforcement learning} have demonstrated that it can aid in the optimization of circuits \cite{fosel2021quantum,ostaszewski2021reinforcement} and accomplish tasks related to QAS.
In the RL settings, the entire circuit is the environment, and the action is to choose a gate from the candidate gate set. The agent is trained to select the appropriate gates using the reward. \cite{ostaszewski2021reinforcement} and \cite{du2020quantum} are regarded as pioneering works in the field of QAS, having employed RL methods to QAS. \cite{kuo2021quantum} tested RL on 2-qubit Bell and 3-qubit GHZ states, while \cite{patel2024curriculum} introduced CRLQAS, achieving chemical accuracy for certain chemical Hamiltonians. \textit{Evolutionary algorithms}, particularly genetic algorithms (GAs) have been used to evolve quantum circuits~\cite{williams1998automated}. Previous studies have shown that GA can effectively evolve simple quantum circuits~\cite{las2016genetic,romero2017quantum,lamata2018quantum,huang2022robust}, often outperforming manually designed circuits. However, these methods are computationally expensive, requiring extensive evaluations of generated circuits to determine their ground-truth performance. This presents a significant bottleneck for scaling QAS in large-scale quantum applications. Our approach mitigates this issue by using layer-wise searching, which reduces extensive search overhead.

\textbf{Performance Evaluation Module in QAS. } The one-shot super-circuit approach \cite{du2022quantum,wang2022quantumNAS} involves sharing gate parameters with sub-circuits following a single training session to expedite the evaluation process. However, optimizing the super-circuit poses significant challenges, and there is often a weak correlation between the performance of sub-circuits with inherited parameters and those optimized through individual training.
Predictor-based QAS \cite{zhang2021neural,he2023gsqas} uses neural networks to predict circuit performance. Training such models requires a dataset of representative circuits with ground-truth evaluations, which is computationally expensive due to the large search space. Additionally, these methods often involve circuit training for performance assessment. It has begun exploring non-neural network approaches \cite{he2024training}. Training-free QAS (TF-QAS)~\cite{he2024training} ranks parameterized quantum circuits (PQCs) using two proxies: the number of paths in the circuit's directed acyclic graph (DAG) and expressivity, defined as the ability to explore the entire Hilbert space. However, TF-QAS remains inefficient in the search module and lacks robustness across different native gate sets and topologies. Meanwhile, the SuperCircuit-based framework quantumNAS exhibits unstable performance on VQE applications due to its fixed layout. In contrast, our method demonstrates greater stability and improved performance.

\textbf{QAS for QML.} Several studies have explored quantum circuit search for quantum machine learning (QML) tasks~\cite{du2022quantum, wang2022quantumNAS, anagolum2024elivagar, wang2022quantumnat}. QuantumSupernet~\cite{du2022quantum} and quantumNAS~\cite{wang2022quantumNAS} build upon the classical Supernet~\cite{pham2018efficient} framework but rely on costly gradient computations. \cite{anagolum2024elivagar} introduce representational capacity as a metric for intra-class similarity and inter-class separation, providing an efficient approach to reducing training overhead in Supercircuit-based QML tasks. However, our work differs as we focus on the VQE task, where their proposed measure is not directly applicable. Additionally, our framework addresses the bottlenecks associated with the Supernet approach.

\textbf{Hardware-aware application implementation.} Many compilation techniques have been developed to minimize circuit transformation costs when deploying quantum applications across different hardware platforms. The mainstream focus has been on superconducting systems, with significant research in key areas such as qubit mapping and routing~\cite{baker2020time, liu2021relaxed, liu2022not, lye2015determining, murali2019noise, nannicini2022optimal, oddi2018greedy, patel2022geyser_atom, smith2021error, das2021adapt, wille2014optimal}, error mitigation~\cite{das2021adapt, das2021jigsaw, patel2021qraft, patel2020disq, smith2021error, stein2023q}, error correction~\cite{wang2024optimizing, yin2025qecc, yin2024surf}, circuit synthesis~\cite{iten2022exact, nam2018automated, xu2022quartz, patel2022quest, younis2021qfast,  peterson2022optimal, weiden2022topology} and pulse-level optimization~\cite{gokhale2020optimized, liang2022pan, meirom2023pansatz, shi2019optimized, zhu2024leveraging} have been proposed. Beyond superconducting devices, optimization efforts in trapped-ion systems primarily focus on minimizing the number of shuttles~\cite{wu2019ilp_iontrap, wu2021tilt, murali2020architecting_iontrap, wu2024boss, zhu2025s}, while in neutral-atom devices, the emphasis is on efficient atom movement~\cite{lin2024reuse, wang2024atomique, bochenneutralatom, hanruiwangQpilot, tan2022qubit}. These techniques complement our method, as they can be integrated to enhance the training performance of our method searched circuits on the target hardware.

\section{Concluding Remarks}\label{sec:conclusion}

In this work, we propose a scalable zero-shot quantum architecture search framework based on the fluctuation of landscape. Our approach uses a layer-wise search strategy complemented by redundancy gate elimination, offering a classical resource efficient solution. We conduct extensive simulations on several well-known Hamiltonians to demonstrate the accuracy and its superior performance compared to previous work. Additionally, we showcase the improvements in handling random Hamiltonians. Furthermore, we perform a large-scale analysis, revealing that our method requires substantially less memory and computational resources than current QAS methods. Our approach successfully handles 50-qubit VQE applications, highlighting its scalability and potential for extending QAS to large-scale applications. 

\textbf{Outlook.} For future work, we aim to integrate our predictor into existing reinforcement learning (RL)-based and evolutionary algorithm-based search modules. This integration could enhance the search process and further improve overall accuracy, enabling more accurate large-scale QAS. Furthermore, we aim to employ noise-adaptive compilation policies for superconducting computers~\cite{murali2019noise, peters2022noise, sharma2023noise, hour2024improving, wang2022quantumNAS} to optimize the search for noisy quantum hardware and we aim to incorporate the evolutionary search and noise-aware performance estimator from \cite{wang2022quantumNAS, wang2022quantumnat} to enhance the noise resilience of our framework.

\section{Acknowledgments}
We would like to thank Tengxiang Lin and Yuqi Li for their insightful discussions.
This work was partially supported by the National Key R\&D Program of China (Grant No.~2024YFE0102500),  the National Natural Science Foundation of China (Grant No.~12447107), the Guangdong Natural Science Foundation (Grant No.~2025A1515012834), the Guangdong Provincial Quantum Science Strategic Initiative (Grant No.~GDZX2403008), the Guangdong Provincial Key Lab of Integrated Communication, Sensing and Computation for Ubiquitous Internet of Things (Grant No.~2023B1212010007), the Quantum Science Center of Guangdong-Hong Kong-Macao Greater Bay Area, and the Education Bureau of Guangzhou Municipality. G. Li was supported by National Natural Science Foundation of China (Grant No. 62402206).

% Use \bibliography{yourbibfile} instead or the References section will not appear in your paper
\bibliography{aaai22}

\begin{thebibliography}{93}
\providecommand{\natexlab}[1]{#1}

\bibitem[{Anagolum et~al.(2024)Anagolum, Alavisamani, Das, Qureshi, and Shi}]{anagolum2024elivagar}
Anagolum, S.; Alavisamani, N.; Das, P.; Qureshi, M.; and Shi, Y. 2024.
\newblock {\'E}liv{\'a}gar: Efficient quantum circuit search for classification.
\newblock In \emph{Proceedings of the 29th ACM International Conference on Architectural Support for Programming Languages and Operating Systems, Volume 2}, 336--353.

\bibitem[{Anschuetz and Kiani(2022)}]{Anschuetz2022}
Anschuetz, E.~R.; and Kiani, B.~T. 2022.
\newblock {Quantum variational algorithms are swamped with traps}.
\newblock \emph{Nature Communications}, 13(1).

\bibitem[{Baker et~al.(2020)Baker, Duckering, Hoover, and Chong}]{baker2020time}
Baker, J.~M.; Duckering, C.; Hoover, A.; and Chong, F.~T. 2020.
\newblock Time-sliced quantum circuit partitioning for modular architectures.
\newblock In \emph{Proceedings of the 17th ACM International Conference on Computing Frontiers}, 98--107.

\bibitem[{Barenco et~al.(1995)Barenco, Bennett, Cleve, DiVincenzo, Margolus, Shor, Sleator, Smolin, and Weinfurter}]{Barenco_1995}
Barenco, A.; Bennett, C.~H.; Cleve, R.; DiVincenzo, D.~P.; Margolus, N.; Shor, P.; Sleator, T.; Smolin, J.~A.; and Weinfurter, H. 1995.
\newblock Elementary gates for quantum computation.
\newblock \emph{Physical Review A}, 52(5): 3457–3467.

\bibitem[{Bergholm et~al.(2022)Bergholm, Izaac, Schuld, Gogolin, Ahmed, Ajith, Alam, Alonso-Linaje, AkashNarayanan, Asadi, Arrazola, Azad, Banning, Blank, Bromley, Cordier, Ceroni, Delgado, Matteo, Dusko, Garg, Guala, Hayes, Hill, Ijaz, Isacsson, Ittah, Jahangiri, Jain, Jiang, Khandelwal, Kottmann, Lang, Lee, Loke, Lowe, McKiernan, Meyer, Montañez-Barrera, Moyard, Niu, O'Riordan, Oud, Panigrahi, Park, Polatajko, Quesada, Roberts, Sá, Schoch, Shi, Shu, Sim, Singh, Strandberg, Soni, Száva, Thabet, Vargas-Hernández, Vincent, Vitucci, Weber, Wierichs, Wiersema, Willmann, Wong, Zhang, and Killoran}]{bergholm2022pennylane}
Bergholm, V.; Izaac, J.; Schuld, M.; Gogolin, C.; Ahmed, S.; Ajith, V.; Alam, M.~S.; Alonso-Linaje, G.; AkashNarayanan, B.; Asadi, A.; Arrazola, J.~M.; Azad, U.; Banning, S.; Blank, C.; Bromley, T.~R.; Cordier, B.~A.; Ceroni, J.; Delgado, A.; Matteo, O.~D.; Dusko, A.; Garg, T.; Guala, D.; Hayes, A.; Hill, R.; Ijaz, A.; Isacsson, T.; Ittah, D.; Jahangiri, S.; Jain, P.; Jiang, E.; Khandelwal, A.; Kottmann, K.; Lang, R.~A.; Lee, C.; Loke, T.; Lowe, A.; McKiernan, K.; Meyer, J.~J.; Montañez-Barrera, J.~A.; Moyard, R.; Niu, Z.; O'Riordan, L.~J.; Oud, S.; Panigrahi, A.; Park, C.-Y.; Polatajko, D.; Quesada, N.; Roberts, C.; Sá, N.; Schoch, I.; Shi, B.; Shu, S.; Sim, S.; Singh, A.; Strandberg, I.; Soni, J.; Száva, A.; Thabet, S.; Vargas-Hernández, R.~A.; Vincent, T.; Vitucci, N.; Weber, M.; Wierichs, D.; Wiersema, R.; Willmann, M.; Wong, V.; Zhang, S.; and Killoran, N. 2022.
\newblock PennyLane: Automatic differentiation of hybrid quantum-classical computations.
\newblock arXiv:1811.04968.

\bibitem[{Bharti et~al.(2022)Bharti, Cervera-Lierta, Kyaw, Haug, Alperin-Lea, Anand, Degroote, Heimonen, Kottmann, Menke, Mok, Sim, Kwek, and Aspuru-Guzik}]{Bharti2022}
Bharti, K.; Cervera-Lierta, A.; Kyaw, T.~H.; Haug, T.; Alperin-Lea, S.; Anand, A.; Degroote, M.; Heimonen, H.; Kottmann, J.~S.; Menke, T.; Mok, W.-K.; Sim, S.; Kwek, L.-C.; and Aspuru-Guzik, A. 2022.
\newblock {Noisy intermediate-scale quantum algorithms}.
\newblock \emph{Reviews of Modern Physics}, 94(1): 015004.

\bibitem[{Bittel and Kliesch(2021)}]{Bittel2021}
Bittel, L.; and Kliesch, M. 2021.
\newblock {Training Variational Quantum Algorithms Is NP-Hard}.
\newblock \emph{Physical Review Letters}, 127(12): 120502.

\bibitem[{Bluvstein et~al.(2024)Bluvstein, Evered, Geim, Li, Zhou, Manovitz, Ebadi, Cain, Kalinowski, Hangleiter, {Bonilla Ataides}, Maskara, Cong, Gao, {Sales Rodriguez}, Karolyshyn, Semeghini, Gullans, Greiner, Vuleti{\'{c}}, and Lukin}]{Bluvstein2024}
Bluvstein, D.; Evered, S.~J.; Geim, A.~A.; Li, S.~H.; Zhou, H.; Manovitz, T.; Ebadi, S.; Cain, M.; Kalinowski, M.; Hangleiter, D.; {Bonilla Ataides}, J.~P.; Maskara, N.; Cong, I.; Gao, X.; {Sales Rodriguez}, P.; Karolyshyn, T.; Semeghini, G.; Gullans, M.~J.; Greiner, M.; Vuleti{\'{c}}, V.; and Lukin, M.~D. 2024.
\newblock {Logical quantum processor based on reconfigurable atom arrays}.
\newblock \emph{Nature}, 626(7997): 58--65.

\bibitem[{Cerezo et~al.(2021)Cerezo, Arrasmith, Babbush, Benjamin, Endo, Fujii, McClean, Mitarai, Yuan, Cincio, and Coles}]{Cerezo2021a}
Cerezo, M.; Arrasmith, A.; Babbush, R.; Benjamin, S.~C.; Endo, S.; Fujii, K.; McClean, J.~R.; Mitarai, K.; Yuan, X.; Cincio, L.; and Coles, P.~J. 2021.
\newblock {Variational quantum algorithms}.
\newblock \emph{Nature Reviews Physics}, 3(9): 625--644.

\bibitem[{Cirac et~al.(2021)Cirac, Pérez-García, Schuch, and Verstraete}]{Cirac_2021}
Cirac, J.~I.; Pérez-García, D.; Schuch, N.; and Verstraete, F. 2021.
\newblock Matrix product states and projected entangled pair states: Concepts, symmetries, theorems.
\newblock \emph{Reviews of Modern Physics}, 93(4).

\bibitem[{Das et~al.(2021)Das, Tannu, Dangwal, and Qureshi}]{das2021adapt}
Das, P.; Tannu, S.; Dangwal, S.; and Qureshi, M. 2021.
\newblock Adapt: Mitigating idling errors in qubits via adaptive dynamical decoupling.
\newblock In \emph{MICRO-54: 54th Annual IEEE/ACM International Symposium on Microarchitecture}, 950--962.

\bibitem[{Das, Tannu, and Qureshi(2021)}]{das2021jigsaw}
Das, P.; Tannu, S.; and Qureshi, M. 2021.
\newblock Jigsaw: Boosting fidelity of nisq programs via measurement subsetting.
\newblock In \emph{MICRO-54: 54th Annual IEEE/ACM International Symposium on Microarchitecture}, 937--949.

\bibitem[{Du et~al.(2020)Du, Huang, You, Hsieh, and Tao}]{du2020quantum}
Du, Y.; Huang, T.; You, S.; Hsieh, M.-H.; and Tao, D. 2020.
\newblock Quantum circuit architecture search: error mitigation and trainability enhancement for variational quantum solvers.
\newblock \emph{arXiv preprint arXiv:2010.10217}.

\bibitem[{Du et~al.(2022)Du, Huang, You, Hsieh, and Tao}]{du2022quantum}
Du, Y.; Huang, T.; You, S.; Hsieh, M.-H.; and Tao, D. 2022.
\newblock Quantum circuit architecture search for variational quantum algorithms.
\newblock \emph{npj Quantum Information}, 8(1): 62.

\bibitem[{Edmonds(1965)}]{edmonds1965paths}
Edmonds, J. 1965.
\newblock Paths, trees, and flowers.
\newblock \emph{Canadian Journal of mathematics}, 17: 449--467.

\bibitem[{Farhi, Goldstone, and Gutmann(2014{\natexlab{a}})}]{Farhi2014}
Farhi, E.; Goldstone, J.; and Gutmann, S. 2014{\natexlab{a}}.
\newblock {A Quantum Approximate Optimization Algorithm}.
\newblock \emph{arXiv:1411.4028}, 1--16.

\bibitem[{Farhi, Goldstone, and Gutmann(2014{\natexlab{b}})}]{farhi2014quantum}
Farhi, E.; Goldstone, J.; and Gutmann, S. 2014{\natexlab{b}}.
\newblock A quantum approximate optimization algorithm.
\newblock \emph{arXiv preprint arXiv:1411.4028}.

\bibitem[{F{\"o}sel et~al.(2021)F{\"o}sel, Niu, Marquardt, and Li}]{fosel2021quantum}
F{\"o}sel, T.; Niu, M.~Y.; Marquardt, F.; and Li, L. 2021.
\newblock Quantum circuit optimization with deep reinforcement learning.
\newblock \emph{arXiv preprint arXiv:2103.07585}.

\bibitem[{Gokhale et~al.(2020)Gokhale, Javadi-Abhari, Earnest, Shi, and Chong}]{gokhale2020optimized}
Gokhale, P.; Javadi-Abhari, A.; Earnest, N.; Shi, Y.; and Chong, F.~T. 2020.
\newblock Optimized quantum compilation for near-term algorithms with openpulse.
\newblock In \emph{2020 53rd Annual IEEE/ACM International Symposium on Microarchitecture (MICRO)}, 186--200. IEEE.

\bibitem[{He et~al.(2023)He, Deng, Zheng, Li, and Situ}]{he2023gsqas}
He, Z.; Deng, M.; Zheng, S.; Li, L.; and Situ, H. 2023.
\newblock Gsqas: graph self-supervised quantum architecture search.
\newblock \emph{Physica A: Statistical Mechanics and its Applications}, 630: 129286.

\bibitem[{He et~al.(2024)He, Deng, Zheng, Li, and Situ}]{he2024training}
He, Z.; Deng, M.; Zheng, S.; Li, L.; and Situ, H. 2024.
\newblock Training-free quantum architecture search.
\newblock In \emph{Proceedings of the AAAI Conference on Artificial Intelligence}, volume~38, 12430--12438.

\bibitem[{Holmes et~al.(2022)Holmes, Sharma, Cerezo, and Coles}]{Holmes2021}
Holmes, Z.; Sharma, K.; Cerezo, M.; and Coles, P.~J. 2022.
\newblock {Connecting Ansatz Expressibility to Gradient Magnitudes and Barren Plateaus}.
\newblock \emph{PRX Quantum}, 3(1): 010313.

\bibitem[{Hour et~al.(2024)Hour, Heng, Go, and Han}]{hour2024improving}
Hour, L.; Heng, S.; Go, M.; and Han, Y. 2024.
\newblock Improving zero-noise extrapolation for quantum-gate error mitigation using a noise-aware folding method.
\newblock \emph{arXiv preprint arXiv:2401.12495}.

\bibitem[{Hu, Choi, and You(2023)}]{Hu2023}
Hu, H.~Y.; Choi, S.; and You, Y.~Z. 2023.
\newblock {Classical shadow tomography with locally scrambled quantum dynamics}.
\newblock \emph{Physical Review Research}, 5(2): 1--26.

\bibitem[{Huang et~al.(2022)Huang, Li, Hou, Wu, Yung, Bayat, and Wang}]{huang2022robust}
Huang, Y.; Li, Q.; Hou, X.; Wu, R.; Yung, M.-H.; Bayat, A.; and Wang, X. 2022.
\newblock Robust resource-efficient quantum variational ansatz through an evolutionary algorithm.
\newblock \emph{Physical Review A}, 105(5): 052414.

\bibitem[{Iten et~al.(2022)Iten, Moyard, Metger, Sutter, and Woerner}]{iten2022exact}
Iten, R.; Moyard, R.; Metger, T.; Sutter, D.; and Woerner, S. 2022.
\newblock Exact and practical pattern matching for quantum circuit optimization.
\newblock \emph{ACM Transactions on Quantum Computing}, 3(1): 1--41.

\bibitem[{Khaneja and Glaser(2000)}]{khaneja2000cartan}
Khaneja, N.; and Glaser, S. 2000.
\newblock Cartan Decomposition of SU($2^n$), Constructive Controllability of Spin systems and Universal Quantum Computing.
\newblock arXiv:quant-ph/0010100.

\bibitem[{Kuo, Fang, and Chen(2021)}]{kuo2021quantum}
Kuo, E.-J.; Fang, Y.-L.~L.; and Chen, S. Y.-C. 2021.
\newblock Quantum architecture search via deep reinforcement learning.
\newblock \emph{arXiv preprint arXiv:2104.07715}.

\bibitem[{Kusyk, Saeed, and Uyar(2021)}]{kusyk2021survey}
Kusyk, J.; Saeed, S.~M.; and Uyar, M.~U. 2021.
\newblock Survey on quantum circuit compilation for noisy intermediate-scale quantum computers: Artificial intelligence to heuristics.
\newblock \emph{IEEE Transactions on Quantum Engineering}, 2: 1--16.

\bibitem[{Lamata et~al.(2018)Lamata, Alvarez-Rodriguez, Martin-Guerrero, Sanz, and Solano}]{lamata2018quantum}
Lamata, L.; Alvarez-Rodriguez, U.; Martin-Guerrero, J.~D.; Sanz, M.; and Solano, E. 2018.
\newblock Quantum autoencoders via quantum adders with genetic algorithms.
\newblock \emph{Quantum Science and Technology}, 4(1): 014007.

\bibitem[{Las~Heras et~al.(2016)Las~Heras, Alvarez-Rodriguez, Solano, and Sanz}]{las2016genetic}
Las~Heras, U.; Alvarez-Rodriguez, U.; Solano, E.; and Sanz, M. 2016.
\newblock Genetic algorithms for digital quantum simulations.
\newblock \emph{Physical review letters}, 116(23): 230504.

\bibitem[{Li et~al.(2024)Li, Hoang, Bhardwaj, Lin, Wang, and Marculescu}]{li2024zero}
Li, G.; Hoang, D.; Bhardwaj, K.; Lin, M.; Wang, Z.; and Marculescu, R. 2024.
\newblock Zero-Shot Neural Architecture Search: Challenges, Solutions, and Opportunities.
\newblock \emph{IEEE Transactions on Pattern Analysis and Machine Intelligence}.

\bibitem[{Liang et~al.(2022)Liang, Cheng, Ren, Wang, Hua, Ding, Chong, Han, Shi, and Qian}]{liang2022pan}
Liang, Z.; Cheng, J.; Ren, H.; Wang, H.; Hua, F.; Ding, Y.; Chong, F.; Han, S.; Shi, Y.; and Qian, X. 2022.
\newblock Pan: Pulse ansatz on nisq machines.
\newblock \emph{CoRR}.

\bibitem[{Lin, Tan, and Cong(2024)}]{lin2024reuse}
Lin, W.-H.; Tan, D.~B.; and Cong, J. 2024.
\newblock Reuse-aware compilation for zoned quantum architectures based on neutral atoms.
\newblock \emph{arXiv preprint arXiv:2411.11784}.

\bibitem[{Liu, Simonyan, and Yang(2018)}]{liu2018darts}
Liu, H.; Simonyan, K.; and Yang, Y. 2018.
\newblock Darts: Differentiable architecture search.
\newblock \emph{arXiv preprint arXiv:1806.09055}.

\bibitem[{Liu, Bello, and Zhou(2021)}]{liu2021relaxed}
Liu, J.; Bello, L.; and Zhou, H. 2021.
\newblock Relaxed peephole optimization: A novel compiler optimization for quantum circuits.
\newblock In \emph{2021 IEEE/ACM International Symposium on Code Generation and Optimization (CGO)}, 301--314. IEEE.

\bibitem[{Liu, Li, and Zhou(2022)}]{liu2022not}
Liu, J.; Li, P.; and Zhou, H. 2022.
\newblock Not all swaps have the same cost: A case for optimization-aware qubit routing.
\newblock In \emph{2022 IEEE International Symposium on High-Performance Computer Architecture (HPCA)}, 709--725. IEEE.

\bibitem[{Lye, Wille, and Drechsler(2015)}]{lye2015determining}
Lye, A.; Wille, R.; and Drechsler, R. 2015.
\newblock Determining the minimal number of swap gates for multi-dimensional nearest neighbor quantum circuits.
\newblock In \emph{The 20th Asia and South Pacific Design Automation Conference}, 178--183. IEEE.

\bibitem[{McClean et~al.(2018)McClean, Boixo, Smelyanskiy, Babbush, and Neven}]{mcclean2018barren}
McClean, J.~R.; Boixo, S.; Smelyanskiy, V.~N.; Babbush, R.; and Neven, H. 2018.
\newblock Barren plateaus in quantum neural network training landscapes.
\newblock \emph{Nature communications}, 9(1): 4812.

\bibitem[{Meirom and Frankel(2023)}]{meirom2023pansatz}
Meirom, D.; and Frankel, S.~H. 2023.
\newblock Pansatz: Pulse-based ansatz for variational quantum algorithms.
\newblock \emph{Frontiers in Quantum Science and Technology}, 2: 1273581.

\bibitem[{Mooney et~al.(2021)Mooney, White, Hill, and Hollenberg}]{Mooney_2021}
Mooney, G.~J.; White, G. A.~L.; Hill, C.~D.; and Hollenberg, L. C.~L. 2021.
\newblock Whole‐Device Entanglement in a 65‐Qubit Superconducting Quantum Computer.
\newblock \emph{Advanced Quantum Technologies}, 4(10).

\bibitem[{Murali et~al.(2019)Murali, Baker, Javadi-Abhari, Chong, and Martonosi}]{murali2019noise}
Murali, P.; Baker, J.~M.; Javadi-Abhari, A.; Chong, F.~T.; and Martonosi, M. 2019.
\newblock Noise-adaptive compiler mappings for noisy intermediate-scale quantum computers.
\newblock In \emph{Proceedings of the twenty-fourth international conference on architectural support for programming languages and operating systems}, 1015--1029.

\bibitem[{Murali et~al.(2020)Murali, Debroy, Brown, and Martonosi}]{murali2020architecting_iontrap}
Murali, P.; Debroy, D.~M.; Brown, K.~R.; and Martonosi, M. 2020.
\newblock Architecting noisy intermediate-scale trapped ion quantum computers.
\newblock In \emph{2020 ACM/IEEE 47th Annual International Symposium on Computer Architecture (ISCA)}, 529--542. IEEE.

\bibitem[{Nam et~al.(2018)Nam, Ross, Su, Childs, and Maslov}]{nam2018automated}
Nam, Y.; Ross, N.~J.; Su, Y.; Childs, A.~M.; and Maslov, D. 2018.
\newblock Automated optimization of large quantum circuits with continuous parameters.
\newblock \emph{npj Quantum Information}, 4(1): 23.

\bibitem[{Nannicini et~al.(2022)Nannicini, Bishop, G{\"u}nl{\"u}k, and Jurcevic}]{nannicini2022optimal}
Nannicini, G.; Bishop, L.~S.; G{\"u}nl{\"u}k, O.; and Jurcevic, P. 2022.
\newblock Optimal qubit assignment and routing via integer programming.
\newblock \emph{ACM Transactions on Quantum Computing}, 4(1): 1--31.

\bibitem[{Oddi and Rasconi(2018)}]{oddi2018greedy}
Oddi, A.; and Rasconi, R. 2018.
\newblock Greedy randomized search for scalable compilation of quantum circuits.
\newblock In \emph{Integration of Constraint Programming, Artificial Intelligence, and Operations Research: 15th International Conference, CPAIOR 2018, Delft, The Netherlands, June 26--29, 2018, Proceedings 15}, 446--461. Springer.

\bibitem[{Ostaszewski et~al.(2021)Ostaszewski, Trenkwalder, Masarczyk, Scerri, and Dunjko}]{ostaszewski2021reinforcement}
Ostaszewski, M.; Trenkwalder, L.~M.; Masarczyk, W.; Scerri, E.; and Dunjko, V. 2021.
\newblock Reinforcement learning for optimization of variational quantum circuit architectures.
\newblock \emph{Advances in Neural Information Processing Systems}, 34: 18182--18194.

\bibitem[{Patel, Silver, and Tiwari(2022)}]{patel2022geyser_atom}
Patel, T.; Silver, D.; and Tiwari, D. 2022.
\newblock Geyser: a compilation framework for quantum computing with neutral atoms.
\newblock In \emph{Proceedings of the 49th Annual International Symposium on Computer Architecture}, 383--395.

\bibitem[{Patel and Tiwari(2020)}]{patel2020disq}
Patel, T.; and Tiwari, D. 2020.
\newblock Disq: a novel quantum output state classification method on ibm quantum computers using openpulse.
\newblock In \emph{Proceedings of the 39th International Conference on Computer-Aided Design}, 1--9.

\bibitem[{Patel and Tiwari(2021)}]{patel2021qraft}
Patel, T.; and Tiwari, D. 2021.
\newblock Qraft: reverse your Quantum circuit and know the correct program output.
\newblock In \emph{Proceedings of the 26th ACM International Conference on Architectural Support for Programming Languages and Operating Systems}, 443--455.

\bibitem[{Patel et~al.(2022)Patel, Younis, Iancu, de~Jong, and Tiwari}]{patel2022quest}
Patel, T.; Younis, E.; Iancu, C.; de~Jong, W.; and Tiwari, D. 2022.
\newblock Quest: systematically approximating quantum circuits for higher output fidelity.
\newblock In \emph{Proceedings of the 27th ACM International Conference on Architectural Support for Programming Languages and Operating Systems}, 514--528.

\bibitem[{Patel et~al.(2024)Patel, Kundu, Ostaszewski, Bonet-Monroig, Dunjko, and Danaci}]{patel2024curriculum}
Patel, Y.~J.; Kundu, A.; Ostaszewski, M.; Bonet-Monroig, X.; Dunjko, V.; and Danaci, O. 2024.
\newblock Curriculum reinforcement learning for quantum architecture search under hardware errors.
\newblock \emph{arXiv preprint arXiv:2402.03500}.

\bibitem[{Peters et~al.(2022)Peters, Shyamsundar, Li, and Perdue}]{peters2022noise}
Peters, E.; Shyamsundar, P.; Li, A.; and Perdue, G. 2022.
\newblock Noise-aware qubit assignment on NISQ hardware using simulated annealing and Loschmidt Echoes.
\newblock \emph{arXiv preprint arXiv:2201.00445}, 10.

\bibitem[{Peterson, Bishop, and Javadi-Abhari(2022)}]{peterson2022optimal}
Peterson, E.~C.; Bishop, L.~S.; and Javadi-Abhari, A. 2022.
\newblock Optimal synthesis into fixed xx interactions.
\newblock \emph{Quantum}, 6: 696.

\bibitem[{Pham et~al.(2018)Pham, Guan, Zoph, Le, and Dean}]{pham2018efficient}
Pham, H.; Guan, M.; Zoph, B.; Le, Q.; and Dean, J. 2018.
\newblock Efficient neural architecture search via parameters sharing.
\newblock In \emph{International conference on machine learning}, 4095--4104. PMLR.

\bibitem[{Preskill(2018)}]{Preskill2018}
Preskill, J. 2018.
\newblock {Quantum Computing in the NISQ era and beyond}.
\newblock \emph{Quantum}, 2: 79.

\bibitem[{Qin et~al.(2020)Qin, Chung, Shi, Vitali, Hubig, Schollw{\"{o}}ck, White, and Zhang}]{Qin2020}
Qin, M.; Chung, C.-M.; Shi, H.; Vitali, E.; Hubig, C.; Schollw{\"{o}}ck, U.; White, S.~R.; and Zhang, S. 2020.
\newblock {Absence of Superconductivity in the Pure Two-Dimensional Hubbard Model}.
\newblock \emph{Physical Review X}, 10(3): 031016.

\bibitem[{Romero, Olson, and Aspuru-Guzik(2017)}]{romero2017quantum}
Romero, J.; Olson, J.~P.; and Aspuru-Guzik, A. 2017.
\newblock Quantum autoencoders for efficient compression of quantum data.
\newblock \emph{Quantum Science and Technology}, 2(4): 045001.

\bibitem[{Sharma and Banerjee(2023)}]{sharma2023noise}
Sharma, A.; and Banerjee, A. 2023.
\newblock Noise-aware Token Swapping for Qubit Routing.
\newblock In \emph{2023 IEEE International Conference on Quantum Computing and Engineering (QCE)}, volume~1, 82--88. IEEE.

\bibitem[{Shi et~al.(2019)Shi, Leung, Gokhale, Rossi, Schuster, Hoffmann, and Chong}]{shi2019optimized}
Shi, Y.; Leung, N.; Gokhale, P.; Rossi, Z.; Schuster, D.~I.; Hoffmann, H.; and Chong, F.~T. 2019.
\newblock Optimized compilation of aggregated instructions for realistic quantum computers.
\newblock In \emph{Proceedings of the Twenty-Fourth International Conference on Architectural Support for Programming Languages and Operating Systems}, 1031--1044.

\bibitem[{Smith et~al.(2021)Smith, Ravi, Murali, Baker, Earnest, Javadi-Abhari, and Chong}]{smith2021error}
Smith, K.~N.; Ravi, G.~S.; Murali, P.; Baker, J.~M.; Earnest, N.; Javadi-Abhari, A.; and Chong, F.~T. 2021.
\newblock Error mitigation in quantum computers through instruction scheduling.
\newblock \emph{arXiv preprint arXiv:2105.01760}.

\bibitem[{Stein et~al.(2023)Stein, Wiebe, Ding, Ang, and Li}]{stein2023q}
Stein, S.; Wiebe, N.; Ding, Y.; Ang, J.; and Li, A. 2023.
\newblock Q-beep: Quantum Bayesian error mitigation employing Poisson modeling over the Hamming spectrum.
\newblock In \emph{Proceedings of the 50th Annual International Symposium on Computer Architecture}, 1--13.

\bibitem[{Tan et~al.(2022)Tan, Bluvstein, Lukin, and Cong}]{tan2022qubit}
Tan, B.; Bluvstein, D.; Lukin, M.~D.; and Cong, J. 2022.
\newblock Qubit mapping for reconfigurable atom arrays.
\newblock In \emph{Proceedings of the 41st IEEE/ACM International Conference on Computer-Aided Design}, 1--9.

\bibitem[{Tan, Lin, and Cong(2025)}]{bochenneutralatom}
Tan, D.~B.; Lin, W.-H.; and Cong, J. 2025.
\newblock Compilation for Dynamically Field-Programmable Qubit Arrays with Efficient and Provably Near-Optimal Scheduling.
\newblock In \emph{Proceedings of the 30th Asia and South Pacific Design Automation Conference}, ASPDAC '25, 921–929. New York, NY, USA: Association for Computing Machinery.
\newblock ISBN 9798400706356.

\bibitem[{Wang et~al.(2022{\natexlab{a}})Wang, Ding, Gu, Lin, Pan, Chong, and Han}]{wang2022quantumNAS}
Wang, H.; Ding, Y.; Gu, J.; Lin, Y.; Pan, D.~Z.; Chong, F.~T.; and Han, S. 2022{\natexlab{a}}.
\newblock Quantumnas: Noise-adaptive search for robust quantum circuits.
\newblock In \emph{2022 IEEE International Symposium on High-Performance Computer Architecture (HPCA)}, 692--708. IEEE.

\bibitem[{Wang et~al.(2022{\natexlab{b}})Wang, Gu, Ding, Li, Chong, Pan, and Han}]{wang2022quantumnat}
Wang, H.; Gu, J.; Ding, Y.; Li, Z.; Chong, F.~T.; Pan, D.~Z.; and Han, S. 2022{\natexlab{b}}.
\newblock Quantumnat: quantum noise-aware training with noise injection, quantization and normalization.
\newblock In \emph{Proceedings of the 59th ACM/IEEE design automation conference}, 1--6.

\bibitem[{Wang et~al.(2024{\natexlab{a}})Wang, Liu, Tan, Liu, Gu, Pan, Cong, Acar, and Han}]{wang2024atomique}
Wang, H.; Liu, P.; Tan, D.~B.; Liu, Y.; Gu, J.; Pan, D.~Z.; Cong, J.; Acar, U.~A.; and Han, S. 2024{\natexlab{a}}.
\newblock Atomique: A quantum compiler for reconfigurable neutral atom arrays.
\newblock In \emph{2024 ACM/IEEE 51st Annual International Symposium on Computer Architecture (ISCA)}, 293--309. IEEE.

\bibitem[{Wang et~al.(2024{\natexlab{b}})Wang, Tan, Liu, Liu, Gu, Cong, and Han}]{hanruiwangQpilot}
Wang, H.; Tan, D.~B.; Liu, P.; Liu, Y.; Gu, J.; Cong, J.; and Han, S. 2024{\natexlab{b}}.
\newblock Q-Pilot: Field Programmable Qubit Array Compilation with Flying Ancillas.
\newblock In \emph{Proceedings of the 61st ACM/IEEE Design Automation Conference}, DAC '24. New York, NY, USA: Association for Computing Machinery.
\newblock ISBN 9798400706011.

\bibitem[{Wang et~al.(2024{\natexlab{c}})Wang, Liu, Stein, Ding, Das, Nair, and Li}]{wang2024optimizing}
Wang, M.; Liu, C.; Stein, S.; Ding, Y.; Das, P.; Nair, P.~J.; and Li, A. 2024{\natexlab{c}}.
\newblock Optimizing FTQC Programs through QEC Transpiler and Architecture Codesign.
\newblock \emph{arXiv preprint arXiv:2412.15434}.

\bibitem[{Weiden et~al.(2022)Weiden, Kalloor, Patel, Younis, Iancu, Kubiatowicz, and Team}]{weiden2022topology}
Weiden, M.; Kalloor, J.; Patel, T.; Younis, E.; Iancu, C.; Kubiatowicz, J.; and Team, B. 2022.
\newblock Topology aware unitary synthesis for scalable quantum circuit optimization.
\newblock In \emph{APS March Meeting Abstracts}, volume 2022, N36--001.

\bibitem[{Wille, Lye, and Drechsler(2014)}]{wille2014optimal}
Wille, R.; Lye, A.; and Drechsler, R. 2014.
\newblock Optimal SWAP gate insertion for nearest neighbor quantum circuits.
\newblock In \emph{2014 19th Asia and South Pacific Design Automation Conference (ASP-DAC)}, 489--494. IEEE.

\bibitem[{Williams and Gray(1998)}]{williams1998automated}
Williams, C.~P.; and Gray, A.~G. 1998.
\newblock Automated design of quantum circuits.
\newblock In \emph{NASA International Conference on Quantum Computing and Quantum Communications}, 113--125. Springer.

\bibitem[{Wu et~al.(2024)Wu, Zhu, Wang, and Wang}]{wu2024boss}
Wu, X.; Zhu, C.; Wang, J.; and Wang, X. 2024.
\newblock BOSS: Blocking algorithm for optimizing shuttling scheduling in Ion Trap.
\newblock \emph{arXiv preprint arXiv:2412.03443}.

\bibitem[{Wu et~al.(2021)Wu, Debroy, Ding, Baker, Alexeev, Brown, and Chong}]{wu2021tilt}
Wu, X.-C.; Debroy, D.~M.; Ding, Y.; Baker, J.~M.; Alexeev, Y.; Brown, K.~R.; and Chong, F.~T. 2021.
\newblock Tilt: Achieving higher fidelity on a trapped-ion linear-tape quantum computing architecture.
\newblock In \emph{2021 IEEE International Symposium on High-Performance Computer Architecture (HPCA)}, 153--166. IEEE.

\bibitem[{Wu et~al.(2019)Wu, Ding, Shi, Alexeev, Finkel, Kim, and Chong}]{wu2019ilp_iontrap}
Wu, X.-C.; Ding, Y.; Shi, Y.; Alexeev, Y.; Finkel, H.; Kim, K.; and Chong, F.~T. 2019.
\newblock ILP-based scheduling for linear-tape model trapped-ion quantum computers.
\newblock In \emph{the Proceedings of The International Conference for High Performance Computing, Networking, Storage, and Analysis, Denver Co}.

\bibitem[{Xu et~al.(2024)Xu, Chung, Qin, Schollw{\"{o}}ck, White, and Zhang}]{Xu2024}
Xu, H.; Chung, C.-M.; Qin, M.; Schollw{\"{o}}ck, U.; White, S.~R.; and Zhang, S. 2024.
\newblock {Coexistence of superconductivity with partially filled stripes in the Hubbard model}.
\newblock \emph{Science}, 384(6696): eadh7691.

\bibitem[{Xu et~al.(2022)Xu, Li, Padon, Lin, Pointing, Hirth, Ma, Palsberg, Aiken, Acar et~al.}]{xu2022quartz}
Xu, M.; Li, Z.; Padon, O.; Lin, S.; Pointing, J.; Hirth, A.; Ma, H.; Palsberg, J.; Aiken, A.; Acar, U.~A.; et~al. 2022.
\newblock Quartz: superoptimization of quantum circuits.
\newblock In \emph{Proceedings of the 43rd ACM SIGPLAN International Conference on Programming Language Design and Implementation}, 625--640.

\bibitem[{Yin et~al.(2024)Yin, Fang, Humble, Li, Shi, and Ding}]{yin2024surf}
Yin, K.; Fang, X.; Humble, T.~S.; Li, A.; Shi, Y.; and Ding, Y. 2024.
\newblock Surf-deformer: Mitigating dynamic defects on surface code via adaptive deformation.
\newblock In \emph{2024 57th IEEE/ACM International Symposium on Microarchitecture (MICRO)}, 750--764. IEEE.

\bibitem[{Yin et~al.(2025)Yin, Zhang, Fang, Shi, Humble, Li, and Ding}]{yin2025qecc}
Yin, K.; Zhang, H.; Fang, X.; Shi, Y.; Humble, T.~S.; Li, A.; and Ding, Y. 2025.
\newblock QECC-Synth: A Layout Synthesizer for Quantum Error Correction Codes on Sparse Architectures.
\newblock In \emph{Proceedings of the 30th ACM International Conference on Architectural Support for Programming Languages and Operating Systems, Volume 1}, 876--890.

\bibitem[{Younis et~al.(2021)Younis, Sen, Yelick, and Iancu}]{younis2021qfast}
Younis, E.; Sen, K.; Yelick, K.; and Iancu, C. 2021.
\newblock Qfast: Conflating search and numerical optimization for scalable quantum circuit synthesis.
\newblock In \emph{2021 IEEE International Conference on Quantum Computing and Engineering (QCE)}, 232--243. IEEE.

\bibitem[{Yu, Zhao, and Wei(2023)}]{yu2023simulating}
Yu, H.; Zhao, Y.; and Wei, T.-C. 2023.
\newblock Simulating large-size quantum spin chains on cloud-based superconducting quantum computers.
\newblock \emph{Physical Review Research}, 5(1): 013183.

\bibitem[{Zhang et~al.(2021{\natexlab{a}})Zhang, Gomes, Yao, Orth, and Iadecola}]{Zhang2021}
Zhang, F.; Gomes, N.; Yao, Y.; Orth, P.~P.; and Iadecola, T. 2021{\natexlab{a}}.
\newblock {Adaptive variational quantum eigensolvers for highly excited states}.
\newblock \emph{Physical Review B}, 104(7): 1--10.

\bibitem[{Zhang, Liu, and Zhang(2024)}]{Zhang2024_k}
Zhang, H.-K.; Liu, S.; and Zhang, S.-X. 2024.
\newblock {Absence of Barren Plateaus in Finite Local-Depth Circuits with Long-Range Entanglement}.
\newblock \emph{Physical Review Letters}, 132(15): 150603.

\bibitem[{Zhang et~al.(2023{\natexlab{a}})Zhang, Zhu, Jing, and Wang}]{Zhang2023}
Zhang, H.-K.; Zhu, C.; Jing, M.; and Wang, X. 2023{\natexlab{a}}.
\newblock {Statistical Analysis of Quantum State Learning Process in Quantum Neural Networks}.
\newblock \emph{Advances in Neural Information Processing Systems}.

\bibitem[{Zhang, Zhu, and Wang(2024)}]{zhang2024predicting}
Zhang, H.-K.; Zhu, C.; and Wang, X. 2024.
\newblock Predicting quantum learnability from landscape fluctuation.
\newblock \emph{arXiv preprint arXiv:2406.11805}.

\bibitem[{Zhang et~al.(2023{\natexlab{b}})Zhang, Allcock, Wan, Liu, Sun, Yu, Yang, Qiu, Ye, Chen et~al.}]{zhang2023tensorcircuit}
Zhang, S.-X.; Allcock, J.; Wan, Z.-Q.; Liu, S.; Sun, J.; Yu, H.; Yang, X.-H.; Qiu, J.; Ye, Z.; Chen, Y.-Q.; et~al. 2023{\natexlab{b}}.
\newblock Tensorcircuit: a quantum software framework for the nisq era.
\newblock \emph{Quantum}, 7: 912.

\bibitem[{Zhang et~al.(2021{\natexlab{b}})Zhang, Hsieh, Zhang, and Yao}]{zhang2021neural}
Zhang, S.-X.; Hsieh, C.-Y.; Zhang, S.; and Yao, H. 2021{\natexlab{b}}.
\newblock Neural predictor based quantum architecture search.
\newblock \emph{Machine Learning: Science and Technology}, 2(4): 045027.

\bibitem[{Zhang et~al.(2022)Zhang, Hsieh, Zhang, and Yao}]{zhang2022differentiable}
Zhang, S.-X.; Hsieh, C.-Y.; Zhang, S.; and Yao, H. 2022.
\newblock Differentiable quantum architecture search.
\newblock \emph{Quantum Science and Technology}, 7(4): 045023.

\bibitem[{Zheng et~al.(2017)Zheng, Chung, Corboz, Ehlers, Qin, Noack, Shi, White, Zhang, and Chan}]{Zheng2017}
Zheng, B.-X.; Chung, C.-M.; Corboz, P.; Ehlers, G.; Qin, M.-P.; Noack, R.~M.; Shi, H.; White, S.~R.; Zhang, S.; and Chan, G. K.-L. 2017.
\newblock {Stripe order in the underdoped region of the two-dimensional Hubbard model}.
\newblock \emph{Science}, 358(6367): 1155--1160.

\bibitem[{Zhong et~al.(2020)Zhong, Wang, Deng, Chen, Peng, Luo, Qin, Wu, Ding, Hu, Hu, Yang, Zhang, Li, Li, Jiang, Gan, Yang, You, Wang, Li, Liu, Lu, and Pan}]{Zhong2020}
Zhong, H.-S.; Wang, H.; Deng, Y.-H.; Chen, M.-C.; Peng, L.-C.; Luo, Y.-H.; Qin, J.; Wu, D.; Ding, X.; Hu, Y.; Hu, P.; Yang, X.-Y.; Zhang, W.-J.; Li, H.; Li, Y.; Jiang, X.; Gan, L.; Yang, G.; You, L.; Wang, Z.; Li, L.; Liu, N.-L.; Lu, C.-Y.; and Pan, J.-W. 2020.
\newblock {Quantum computational advantage using photons}.
\newblock \emph{Science}, 370(6523): 1460--1463.

\bibitem[{Zhou et~al.(2020)Zhou, Wang, Choi, Pichler, and Lukin}]{Zhou2020}
Zhou, L.; Wang, S.-T.; Choi, S.; Pichler, H.; and Lukin, M.~D. 2020.
\newblock {Quantum Approximate Optimization Algorithm: Performance, Mechanism, and Implementation on Near-Term Devices}.
\newblock \emph{Physical Review X}, 10(2): 021067.

\bibitem[{Zhu et~al.(2025)Zhu, Wu, Wang, and Wang}]{zhu2025s}
Zhu, C.; Wu, X.; Wang, J.; and Wang, X. 2025.
\newblock S-SYNC: Shuttle and Swap Co-Optimization in Quantum Charge-Coupled Devices.
\newblock \emph{arXiv preprint arXiv:2505.01316}.

\bibitem[{Zhu et~al.(2024)Zhu, Cheng, Li, Zhou, Ding, and Liang}]{zhu2024leveraging}
Zhu, Y.; Cheng, J.; Li, B.; Zhou, Y.; Ding, Y.; and Liang, Z. 2024.
\newblock Leveraging Hardware Power through Optimal Pulse Profiling for Each Qubit Pair.
\newblock \emph{arXiv preprint arXiv:2411.19308}.

\end{thebibliography}

\end{document}